\documentclass{PoS}
\usepackage{latexsym,amssymb,pifont,amsmath,xspace,amsmath,amsfonts,amsthm}
\usepackage{graphicx}

\title{Quantum isolated horizons and black hole entropy}

\ShortTitle{Quantum isolated horizons and black hole entropy}

\author{\speaker{J. Fernando Barbero G.}\\
        Instituto de Estructura de la Materia, CSIC, Serrano 123, 28006 Madrid, Spain\\
        E-mail: \email{fbarbero@iem.cfmac.csic.es}}

\author{Jerzy Lewandowski\\
        Institute of Theoretical Physics, Warsaw University, ul. Ho\.{z}a 69, 00-681 Warsaw, Poland.\\
        E-mail: \email{Jerzy.Lewandowski@fuw.edu.pl}}

\author{Eduardo J. S. Villase\~nor\\
        Instituto Gregorio Mill\'an, Grupo de Modelizaci\'on y Simulaci\'on Num\'erica, Universidad Carlos III de Madrid, Avda. de la Universidad 30, 28911 Legan\'es, Spain\\
        Instituto de Estructura de la Materia, CSIC, Serrano 123, 28006 Madrid, Spain\\
        E-mail: \email{ejsanche@math.uc3m.es}}

\abstract{We give a short introduction to the approaches currently used to describe black holes in loop quantum gravity. We will concentrate on the classical issues related to the modeling of black holes as isolated horizons, give a short discussion of their canonical quantization by using loop quantum gravity techniques, and a description of the combinatorial methods necessary to solve the counting problems involved in the computation of the entropy.}

\FullConference{3rd Quantum Gravity and Quantum Geometry School\\
		February 28 - March 13, 2011\\
		Zakopane, Poland}

\begin{document}

\section{Introduction}

One of the main motivations of any quantum gravity theory is to gain an understanding of the quantum behavior of black holes (BH's), explain in fundamental terms the origin of their entropy and discuss other quantum phenomena such as Hawking radiation. Both string inspired models and loop quantum gravity (LQG) can be used to model quantum (or semiclassical) black holes, describe the microscopic degrees of freedom responsible for their entropy and explain the proportionality between entropy and area (the Bekenstein-Hawking law).

In the string theory side the pioneering work of Strominger and Vafa \cite{Strominger} showed that, in the case of extremal black holes carrying certain types of charges, the statistical entropy can be obtained by counting microscopic BPS states corresponding to the same charges. For large values of the charges the horizon area of the black hole and the number of microstates are proportional and the constant of proportionality is, precisely, 1/4 (in geometric units) in perfect agreement with the area law. For other non-extremal black holes the complementarity principle of Horowitz and Polchinski \cite{Horowitz} explains the proportionality between area and entropy though it does not say anything with regard to the proportionality constant. These days a lot of attention \cite{Mandal} is being payed to precision studies and the comparison of subleading contributions to the statistical entropy for large areas with the generalizations of the Bekenstein-Hawking law suggested by Wald \cite{Wald:1993nt,Iyer:1994ys}.

A somewhat parallel development has taken place in the loop quantum gravity approach to the study of quantum black holes. The first attempts in this direction go back to the work of Smolin \cite{Smolin:1995vq} where he suggested to model quantum black holes as inner spacetime boundaries and highlighted the role of Chern-Simons theory to understand the quantization of the horizon degrees of freedom. Other interesting suggestions by Krasnov and Rovelli followed soon after \cite{Krasnov:1996tb,Krasnov:1996wc,Rovelli:1996dv,Rovelli:1996ti}. A synthesis of the foregoing ideas and the development of the modeling of BH's in LQG by using isolated horizons appears in
\cite{Ashtekar:1997yu,Ashtekar:2000eq}. The starting point of this approach is the Ashtekar Hamiltonian formulation of  spacetimes with isolated horizons as inner boundaries \cite{Ashtekar:1999wa}. By quantizing the resulting model in a Hilbert space obtained by tensoring a Chern-Simons boundary Hilbert space and a standard LQG bulk Hilbert space, it is possible to address the problem of describing the statistical BH entropy in LQG. A practical rephrasing of the combinatorial problem associated with the computation of the entropy was given by Domaga{\l}a and Lewandowski \cite{Domagala:2004jt}. They also pointed out that all the quantum area states must be taken into account to get the quantum horizon entropy and not only the lowest excitations,  as had been stated in \cite{Ashtekar:2000eq} and considered true for several years. Their result affected the value of the quantum of area predicted by LQG. That value was calculated by Meissner \cite{Meissner:2004ju} as well as and the asymptotic behavior of the statistical entropy as a function of the area leading to the Bekenstein-Hawking law.

The era of precision counting in LQG was ushered by Corichi, D\'{\i}az Polo and F.-Borja in \cite{Corichi:2006bs, Corichi:2006wn}. Their main discovery was an effective quantization of the black hole entropy for small areas. Despite the fact that the area spectrum in LQG has a complicated structure (and, in particular, is not equally spaced) the statistical entropy shows a staircase structure with regular steps of constant area width. The first attempts to use combinatorial methods to explain this behavior of the entropy --in a somewhat simplified setting-- appear in the works of Sahlmann \cite{Sahlmann:2007jt,Sahlmann:2007zp} where, in particular, he explored the use of generating functions. In fact, the combinatorial problems are such that it is possible to solve them \textit{exactly} by using number theoretic methods \cite{Agullo:2008yv}, generating functions \cite{BarberoG.:2008ue}, group theory methods, Laplace transforms and other related ideas. A detailed account of this approaches can be found in \cite{Agullo:2010zz}. The very existence of the staircase behavior of the entropy and its eventual disappearance in the large area limit can also be studied by using the same ideas \cite{FernandoBarbero:2011kb}. An important part of the present paper will be devoted to describe them.  An interesting side result of the previous framework is the possibility of obtaining the exact partition functions (in the so called area canonical ensemble). By using them it is straightforward to obtain the ``true'' (thermodynamic)  entropy (as opposed to the statistical or counting entropy) and its asymptotic behavior as a function of the area \cite{G.:2011zr}. The fact that the subleading corrections of the statistical and thermodynamical entropies differ may be important to understand the stability of black holes. This is so because the overall stability of thermal systems is associated to the concavity of the entropy.

To end this introduction we would like to mention the recent development of an explicit $SU(2)$ formulation \cite{Engle:2009vc, Engle:2010kt, Engle:2011vf} that relies on covariant Hamiltonian methods and avoids the partial gauge fixing of the standard approach \cite{Ashtekar:1997yu,Ashtekar:2000eq}. This recent work puts in a rigorous footing some older proposals by Smolin \cite{Smolin:1995vq}, Krasnov \cite{Krasnov:1997yt}, and Kaul and Majumdar \cite{Kaul:1998xv,Kaul:2000kf} where the authors suggested that the quantum black hole degrees of freedom could be described with the help of a $SU(2)$ Chern-Simons theory. An important difference between these two points of view concerns the treatment of the quantum boundary conditions.

The combinatorial problems that have to be solved to compute the entropy differ in some details but can be dealt with by the aforementioned methods. This difference percolates to the physics described by each model, in particular with regard to the subdominant contributions to the asymptotic behavior of the entropy as a function of the area.

\section{Weakly isolated horizons and canonical gravity}

\subsection{Non-expanding and weakly isolated horizons}
Consider a $4$-dimensional spacetime $(M,g)$, where $M$ is a 4-manifold and $g$ is a Lorentzian metric tensor with signature $(-,+,+,+)$. A 3-dimensional hypersurface ${\cal N}\subset M$ is called a \textit{null surface},  whenever the pullback\footnote{Throughout this paper, given an arbitrary covariant tensor $T$ in an arbitrary manifold $M$ and an arbitrary submanifold $N$, the pullback of $T$ to $N$ via the embedding $N\hookrightarrow M$ will be denoted by $T^{(N)}$.}  $g^{({\cal N})}$of the spacetime metric onto ${\cal N}$ is degenerate. This means that at every point of ${\cal N}$ there exists a  non-zero vector $\ell$ tangent to ${\cal N}$ and such that
$$\ell^a g^{({\cal N})}_{ab}X^b= 0,\ \ \ \ \ $$
for every vector $X$ tangent to ${\cal N}$. Being normal to $\mathcal{N}$ the vector field $\ell$ is twist free. It is generally true that the curves tangent to the degenerate directions of $g$ are (unparametrized) null geodesics in $M$ (see, for example, \cite{Kupeli}). Under mild topological conditions \cite{Kupeli} (natural from a physical point of view) the null 3-surface can be thought of as the world-3-surface of a 2-dimensional spacelike surface $S$ each point of which travels in $M$ at the ``speed of light'' in the direction orthogonal to $S$. A spacelike section $\tilde{S}$ of ${\cal N}$ can be thought of as an embedding of the abstract 2-surface $S$ corresponding to a given instant of time. Suppose that the 2-surface $S$ is a sphere and its embedding $\tilde{S}$ is the boundary of a region of $M$ diffeomorphic to a space-like 3-ball. Then, the null world-3-surface ${\cal N}$ is a history of a 2-surface $S$ blowing up or shrinking at the speed of light. In general, the 2-metric induced on $S$ depends on the embedding $\tilde{S}$.

The null surfaces considered in this paper will not be generic because we will add more structure to them. Let us start by considering the expansion of a null vector field $\ell$ tangent to ${\cal N}$ defined as follows
\begin{equation}
\theta_{(\ell)}:= g^{({\cal N})ab}{\cal L}_{\ell}g^{({\cal N})}_{ab}\,,
\end{equation}
where $g^{({\cal N})ab}$ is \textit{any} tensor such that
$$g^{({\cal N})}_{ab}=g^{({\cal N})}_{aa'}g^{({\cal N}) a'b'}g^{({\cal N})}_{b'b} \ .$$
On a null surface the vector field $\ell$ is defined up to local scalings $\ell\mapsto \ell'=f\ell$ and the expansion  transforms as
$\theta_{(\ell)}\mapsto \theta_{(\ell')}=f\theta_{(\ell)}$. Hence, the value of the expansion \textit{of a null surface} at a point $x\in {\cal N}$ is not well defined --it depends on the null normal $\ell$-- unless it is zero because
$$ \theta_{(\ell)}(x)\ =\ 0\ \Leftrightarrow \theta_{(\ell')}(x)\ =\ 0.$$
If $\theta_{(\ell)}(x) = 0$ for all $x\in{\cal N}$, we will say that ${\cal N}$ is a {\it non-expanding null surface}.

For every non-expanding null surface ${\cal N}$, the following physically reasonable inequality
\begin{equation} R_{ab}\ell^a\ell^b\ \ge\ 0\,,\label{Rab}
\end{equation}
involving the Ricci tensor $R_{ab}$, may be assumed to hold at every $x\in {\cal N}$ and for any null vector $\ell$ tangent to ${\cal N}$ at $x$. If this is the case, a geometric identity called the Raychaudhuri equation implies\footnote{In a non-expanding null surface, any null vector field $\ell$ is twist-free and expansion-free and, hence, the Raychaudhuri equation implies that it is also shear-free \cite{AFK}.} that
\begin{equation}\label{lielq} {\cal L}_{\ell}g^{({\cal N})}_{ab}\ =\ 0\,. \end{equation}
This property independent of the choice of the null
vector field $\ell$  tangent to ${\cal N}$ in the sense that if it holds for one choice, then it
is true for all of them. According to the Einstein equations, the assumption (\ref{Rab}) about the Ricci tensor is equivalent to
$$ R_{ab}\ell^a\ell^b\ =\ 8\pi G T_{ab}\ell^a\ell^b\ \ge\ 0.$$
This is the matter energy-density positivity, a physically justified assumption, in fact, as a consequence of the Raychaudhuri equation, the stronger condition $T_{ab}\ell^a\ell^b=0$ holds. An example of a non-expanding null surface for which (\ref{Rab}) is true is a null plane in Minkowski spacetime, therefore, there is nothing necessarily exotic about this type of null surfaces.
What makes a non-expanding null surface ${\cal N}$ a {\it non-expanding horizon} (NEH) are, in addition to (\ref{Rab}), the following topological and geometrical assumptions: \\

$\bullet$ There exists a  diffeomorphism  $S\times(0,1)\rightarrow {\cal N}$, where $S$  is a 2-sphere.

$\bullet$  The intervals $\{x\}\times (0,1)$ correspond, via this diffeomorphism, to  null geodesics in ${\cal N}$.

$\bullet$  Finally, each global section of $S\times(0,1)$ corresponds to a space-like 2-surface in ${\cal N}$.\\

In summary, a NEH can be thought of as a 2-sphere $S$ which, on one hand, is blowing up  or shrinking at the speed of light but, on the other, satisfies that its intrinsic geometry ``does not change'' in the sense that it is invariant under the action of the diffeomorphisms generated by the null normals.

To finally arrive at the definition of a weakly  isolated horizon we will endow a non-expanding horizons ${\cal N}$ with more structure, namely, we will reduce the arbitrariness in the choice of the null vector field $\ell$ tangent to ${\cal N}$ by fixing a particular one (modulo constant rescalings). To this end, let us recall that the following identity \cite{Kupeli} holds on every null surface\footnote{It is important here that the codimension is one.}
\begin{equation}
\nabla_{\ell}\ell\ =\ \kappa\ell,
\end{equation}
where $\kappa:\ {\cal N}\ \rightarrow\ \mathbb{R}$. This means that, in particular, one could choose $\ell$ in such a way that its integral curves are  affinely parametrized geodesics and, then, $\kappa$ would vanish everywhere.  Instead, we  assume a weaker condition, namely
\begin{equation}\label{kappaconst}
\kappa\ =\ {\rm const}.
\end{equation}
Notice, however, that given a null surface there is still a large family of null vector fields
which satisfy equation (\ref{kappaconst}).

A {\it weakly isolated horizon} (WIH) is a pair $({\cal N},[\ell])$, where
${\cal N}$ is a non-expanding horizon and $[\ell]$ the class (modulo constant rescalings and represented by $\ell$) of null vector field tangents to ${\cal N}$, such that (\ref{kappaconst}) holds. Notice, however, that the value of $\kappa$ is not defined on the class $[\ell]$ and, hence, it is not well define for a WIH.

An interesting and physically relevant example of WIH is provided by the Kerr-Newman spacetimes (including Schwarzschild) for which the event horizon is, in fact, a non-expanding horizon. In each of those cases there is a Killing vector field $\xi$ such that its restriction $\ell$ to the horizon is null and tangent to it. This Killing vector is defined up to a constant rescaling, however, this ambiguity can be removed by imposing a normalization condition at the asymptotically flat region of the spacetime. In addition to this, there is a geometric identity, called ``the $0$th law of black hole thermodynamics'' (see, for example, \cite{Wald}), which implies that for such a Killing vector field $\kappa$ is constant (for this reason $\kappa$ is called the {\it surface gravity}).
Indeed, for every null vector field $\ell$ tangent to a null surface we have
$$ \nabla_a\ell^b\ =\ \omega_a\ell^b, $$
and the following geometric identity holds:
$$ d\kappa\ =\ {\cal L}_\ell\omega\,. $$
Owing to the fact that $\ell$ is the restriction of a spacetime Killing vector field to a null surface, the right hand side to the previous equation has to be zero. Hence $\kappa$ is constant and then $[\ell]$ defines on the black hole event horizon of each Kerr-Newman spacetime the structure of weakly-isolated horizon.

In conclusion, the concept of a weakly-isolated horizon is a \textit{quasi-local} generalization of the definition of a black hole event horizon. Actually, the only global assumption is the topology of the space-like sections of non-expanding horizons. Almost all the theorems concerning the black holes admit quasilocal generalizations to the type of horizons introduced in this section \cite{LP1,LP2,LP3,LP4}.

\subsection{The symplectic 2-form in the space of spacetimes}
\noindent Given a spacetime $(M,g)$ consider a local co-frame $e^I$, with $I=1,\ldots,4$, such that $g\ =\ \eta_{IJ}e^I\otimes e^J$ in a given chart. The constant matrix $\eta_{IJ}$ and its inverse will be used below to lower and raise indices, respectively.
In our paper we will choose the non-zero components of $\eta_{IJ}$ to be
$$\eta_{11}\ =\ \eta_{22}\ =\ -\eta_{34}\ = \ -\eta_{43}\ =\ 1.$$
Given a co-frame $e^I$ we define the corresponding (local) frame connection, that is, the matrix $(\Gamma^I{}_J)$, $I,J=1,\ldots,4$, whose entries are differential 1-forms satisfying
\begin{equation}\label{Gamma}
de^I+\Gamma^I{}_J\wedge e^J\ =\ 0, \ \ \ \ \ \ \Gamma_{IJ}+\Gamma_{JI}\ =\ 0.
\end{equation}
Suppose now that $e^I$ is a solution of the vacuum Einstein equations, and consider a 1-dimensional family of solutions $e^I(s)$ parametrized by $s$, such that $e^I(0)=e^I$. Let us denote
\begin{equation}
\delta e^I\ :=\ \frac{d}{ds}\Big|_{s=0} e^I(s). \label{delta}
\end{equation}
Therefore  $\delta e^I$, to which we will refer to as a {\it variation of the co-frame},
is a vector tangent to the space of solutions to the Einstein equations at the point $e^I$.  For every function $f$ on this space we define the operator
$$\delta f(e^I)\ :=\  \frac{d}{ds}\Big|_{s=0}f(e^I(s))\,.$$
For example, for every value of $s$ we have the frame connection $\Gamma^I{}_J(s)$ corresponding to the co-frame forms $e^I(s)$ and, according to the previous definition, we have
\begin{equation}
\delta \Gamma^I{}_J\ =\ \frac{d}{ds}\Big|_{s=0}{\Gamma^I{}_{J}}(s).
\end{equation}
Given two curves in the space of co-frames with tangent vectors $\delta_1e^I$ and $\delta_2e^I$  at the same point $e^I$, we define the following 3-form in $M$
\begin{equation}\label{current}
\frac{1}{2}\epsilon_{IJKL}\delta_{[1}(e^I\wedge e^J)\wedge \delta_{2]}\Gamma^{KL}:=\frac{1}{2}\epsilon_{IJKL}\left(\delta_1(e^I\wedge e^J)\wedge \delta_2\Gamma^{IJ}\ -\ \delta_2(e^I\wedge e^J)\wedge \delta_1\Gamma^{IJ}\right).
\end{equation}
This 3-form is closed as a consequence of the vacuum\footnote{This is also true for a non-vanishing cosmological constant.} Einstein equations, that is
\begin{equation}
d\left(\frac{1}{2}\epsilon_{IJKL}\delta_{[1}(e^I\wedge e^J)\wedge \delta_{2]}\Gamma^{KL}\right)\ =\ 0
\end{equation}
because
\begin{equation}
\frac{1}{2}\epsilon_{IJKL}e^J\wedge (d\Gamma^{KL}+\Gamma^K{}_M\wedge\Gamma^{ML})\ =\ 0\,.
\end{equation}
The same is true for each element of the family of 3-forms
$$
\frac{1}{2}\epsilon_{IJKL}\delta_{[1}(e^I\wedge e^J)\wedge \delta_{2]}\Gamma^{KL}\ -\ \frac{1}{\gamma}\delta_{[1}(e^I\wedge e^J)\wedge \delta_{2]}\Gamma_{IJ}
$$
labeled by the Immirzi parameter $\gamma$, namely
\begin{equation}
d\left(\frac{1}{2}\epsilon_{IJKL}\delta_{[1}(e^I\wedge e^J)\wedge \delta_{2]}\Gamma^{KL}\ -\ \frac{1}{\gamma}\delta_{[1}(e^I\wedge e^J)\wedge \delta_{2]}\Gamma_{IJ}\right)\ =\ 0\,.
\end{equation}
Therefore, if a spacetime $(M,g)$ satisfies the vacuum Einstein equations, and we are given a family of 3-surfaces $\Sigma$ such that every pair $\Sigma_1$ and $\Sigma_2$ of the family define the boundary of a 4-dimensional submanifold of $M$, then the integral
\begin{equation}\label{integral}
\Omega(\delta_1,\delta_2)\ :=\ \frac{1}{8\pi G}\int_\Sigma \left(\frac{1}{2}\epsilon_{IJKL}\delta_{[1}(e^I\wedge e^J)\wedge \delta_{2]}\Gamma^{KL}\ -\ \frac{1}{\gamma}\delta_{[1}(e^I\wedge e^J)\wedge \delta_{2]}\Gamma_{IJ}\right)
\end{equation}
is independent of the choice of the surface $\Sigma$. Here and in the following we relax the notation in the usual way and denote the tangent vectors $\delta e^I$ simply by $\delta$.

In a more general case, relevant for black holes, we consider a region $M_{\rm ex}$ of spacetime $M$ bounded on one side by a causal 3-surface ${\cal N}$ and a family of 3-surfaces $\Sigma$ such that:
\begin{itemize}
\item Every pair $\Sigma_1$ and $\Sigma_2$, defines a 4-dimensional region of the spacetime bounded by $\Sigma_1$, $\Sigma_2$ {\it and} a segment of the surface ${\cal N}$ contained between the 2-surfaces $\Sigma_1\cap {\cal N}$ and $\Sigma_2\cap {\cal N}$.
\item For every pair of tangent vectors $\delta_1,\delta_2$ there exists a 2-form $\alpha(\delta_1,\delta_2)$ defined on ${\cal N}$ such that
\begin{equation}\label{closedonN}
\left(\frac{1}{2}\epsilon_{IJKL}\delta_{[1}(e^I\wedge e^J)\wedge \delta_{2]}\Gamma^{KL}\ -\ \frac{1}{\gamma}\delta_{[1}(e^I\wedge e^J)\wedge \delta_{2]}\Gamma_{IJ}\right)^{({\cal N})}\ =\ d\alpha(\delta_1,\delta_2)\,.
\end{equation}
\end{itemize}
Under these conditions the following generalization of (\ref{integral})
\begin{eqnarray}\label{integralN}
\Omega(\delta_1,\delta_2)\ &:=&\ \frac{1}{8\pi G}\int_{\Sigma\cap {\cal N}}\alpha(\delta_1,\delta_2) \\ &+&\ \frac{1}{8\pi G}\int_\Sigma \left(\frac{1}{2}\epsilon_{IJKL}\delta_{[1}(e^I\wedge e^J)\wedge \delta_{2]}\Gamma^{KL}\ -\ \frac{1}{\gamma}\delta_{[1}(e^I\wedge e^J)\wedge \delta_{2]}\Gamma_{IJ}\right)\nonumber
\end{eqnarray}
is independent of $\Sigma$. In both cases, (\ref{integral}) or (\ref{integralN}), the map
$(\delta_1,\delta_2)\ \mapsto\ \Omega(\delta_1,\delta_2)$ defines  a bi-linear, anti-symmetric form in the tangent space at $e^I$ to the space of solutions of the Einstein equations (on $M$ or, respectively, $M_{\rm ex}$). By extending this procedure to arbitrary points of the solution space we can define the differential 2-form
$e^I\ \mapsto\ \Omega$, the (pre)-symplectic form on the space of solutions of the Einstein equations.

This symplectic form extends naturally to the space of pairs $(e^I,\Gamma^I{}_J)$ where the 1-forms $\Gamma^I{}_J$  on $M$ are independent of $e^I$ as configuration variables and satisfy the metricity condition $\Gamma_{IJ}\ =\ -\Gamma_{JI}$. Notice that in this first order formalism the structure equation enforcing the torsion-less character of  $\Gamma^I{}_J$ appears as an additional field equation.

To end this subsection we want to point out that even though the 3-form used to define $\Omega$ has been introduced locally it is actually well defined globally. Indeed, if we pick an atlas on $M$, and consider points that belong to two charts, the respective co-frames are related to each other by point dependent $SO(1,3)$ transformations
$$ e^{\,\prime\,\,I}\ =\ \Lambda^I{}_Ke^K,\ \ \ \Gamma^{\,\prime\,\,I}{}_J\ =\ \Lambda^I{}_K\Gamma^K{}_L(\Lambda^{-1})^L{}_J +
\Lambda^I{}_Kd(\Lambda^{-1})^K{}_J. $$
The integrands in (\ref{integral}) and (\ref{integralN}) are, nevertheless, invariant with respect to these and the differential 3-form appearing in the integrand is defined \textit{globally} on $M$.

\subsection{The symplectic form and the weakly isolated horizons}\label{setup}
Let us consider a foliation of $M_{\rm ex}$ by Cauchy surfaces $\Sigma$ intersecting a (segment of a) WIH $({\cal N},[\ell])$. We will calculate now the surface term over $\Sigma\cap{\cal N}$ of the integral (\ref{integralN}).  We will do that in a ``gauge'', that we introduce by choosing a co-frame $e^I$ in such a way that its dual tangent frame satisfies the following conditions:
\begin{enumerate}
\item[\textbf{GF1.}] The vector field dual to $e^4$ coincides with a null normal belonging to the class $[\ell]$, i.e. $ e_4 \ =\ \ell$.
\item[\textbf{GF2.}] The vector fields $e_1$ and $e_2$ are tangent to a foliation of ${\cal N}$ by spacelike sections.
\item[\textbf{GF3.}] The foliation is preserved by the flow of $\ell$.
\end{enumerate}
We will also introduce a function $v:{\cal N}\rightarrow \mathbb{R}$, constant on the leaves of the foliation, and such that
\begin{equation}
\ell^a (dv)_a\ =\ 1\,,
\end{equation}
and asume that the horizon ${\cal N}$ is bounded by the sections $S_{v_0}$ and $S_{v_1}$, where
$S_{v}$ denotes the leaf of the foliation (a section of ${\cal N}$)
defined by a level set of $v$.

The spherical topology of the horizon sections will be relevant in the following, therefore we need to pay attention to the global features of the fields. In particular, there is no obstruction to define the co-frame 1-forms $e^4$ and $e^3$ globally on ${\cal N}$ and such that the pull back $ {e^3}^{({\cal N})}\ =\ 0$. However, the 1-forms $e^1$ and $e^2$ (that can be used to write down the metric tensor induced on each $S_v$) can not be defined globally on ${\cal N}$ (remember that the two sphere cannot be parallelized). In this case it is, nonetheless, possible to use two charts in such a way that the corresponding frames are related by an $SO(1,3)$ rotation in the overlapping region. These two charts can be chosen in such a way that each of them contains the entire null geodesic in the segment ${\cal N}$ (the integral curves of $\ell$). The property (\ref{lielq}) in terms of our frame reads
\begin{equation}{\cal L}_\ell g^{({\cal N})}={\cal L}_\ell (e^1\otimes e^1 + e^2\otimes e^2)^{({\cal N})}\ =\ 0\,,
\label{lielq2}
\end{equation}
and the metric induced on each section $S_{v}$ is
$$ g^{(S_{v})}\ =\  (e^1\otimes e^1 + e^2\otimes e^2)^{({S_{v}})}.$$
Given a WIH, equation (\ref{lielq2}) tells us that all the sections are naturally isometric to each other and, in particular,  have the same  area
$$a\ =\ \int_{S_{v}}e^1\wedge e^2$$
irrespectively of the value of $v$.

Finally, the connection 1-forms $\Gamma^I{}_J$ must satisfy (\ref{Gamma}). We assume  that the variations $\delta_1$ and $\delta_2$  of the co-frame in (\ref{integral}) preserve the above gauge fixing conditions. As a consequence of (\ref{lielq}), the property (\ref{lielq2}) and the gauge choice introduced above, several components of the connection one forms $\Gamma^I{}_J$ vanish and, hence, the  left hand side of (\ref{closedonN}) becomes
\begin{eqnarray}
& & \left(\frac{1}{2}\epsilon_{IJKL}\delta_{[1}(e^I\wedge e^J)\wedge \delta_{2]}\Gamma^{KL}\ -\ \frac{1}{\gamma}\delta_{[1}(e^I\wedge e^J)\wedge \delta_{2]}\Gamma_{IJ}\right)^{({\cal N})}\hspace{5cm}\nonumber\\
& & \hspace{5cm}
 = \ \left(\delta_{[1}(e^1\wedge e^2)\wedge \delta_{2]}\Gamma^{34}\ -\ \frac{1}{\gamma}\delta_{[1}(e^1\wedge e^2)\wedge \delta_{2]}\Gamma_{12}\right)^{({\cal N})}\hspace{-.5cm}.\label{Omega}
\end{eqnarray}
Before proceeding, we will make two remarks.

$\bullet$ The first is that if we restrict ourselves to WIH's such that
\begin{equation}
\ell^a\delta\Gamma^{34}{}_a\ =\ \ell^a\delta\Gamma^{12}{}_a\ =\ 0,
\end{equation}
then the boundary term of the symplectic structure in (\ref{integralN}), that is, the integral along $\Sigma\cap {\cal N}$, is identically zero. In fact, in the case of a non-expanding horizon the previous conditions can be satisfied by performing suitable Lorentz rotations of the co-frame.  Indeed, it is enough to
enlarge the list of the gauge fixing conditions with the addition of the conditions
\begin{equation}\label{stronggauge} \kappa=0, \ \ \ \ \ \ \ \ \  {\cal L}_\ell e^1\ =\ 0\ =\ {\cal L}_\ell e^2\,,\end{equation}
that can be satisfied on every non-expanding horizon by a suitable choice of the null vector field $\ell$ (demanding it to be the tangent to an affinely parametrized null geodesic) and  obtaining the co-frame elements $e^1$ and $e^2$ by Lie dragging in the direction of $\ell$ (recall equation (\ref{lielq2})).

$\bullet$ The second remark is that, in the classical mechanics of WIH \cite{LP3,ABF,ABL1}, the first condition in (\ref{stronggauge}) cannot be imposed because $\ell$ is not fixed, that is, we admit weakly isolated horizons of all the non-zero constant values of $\kappa$. In that case it is easy to calculate that (\ref{closedonN}) holds with
\begin{equation}\label{Omegamechanics}
\alpha(\delta_1,\delta_2)\ =\ 2 \delta_{[1}(e^1\wedge e^2)^{({\cal N})} \delta_{2]}(\kappa)v.
\end{equation}

From this introductory example of the Mechanics of WIH, we learn two
lessons. The first is that in this case the
degrees of freedom which contribute to the symplectic form are pure gauge because
they correspond to the freedom of rescaling a null vector field tangent to
the horizon
$$ \ell\ \mapsto f\ell$$
by functions which preserve the condition $\kappa\ =\ {\rm const}$. The second lesson is that in this case the boundary term of the symplectic structure --the first term in (\ref{integralN})-- takes the following form
\begin{equation}\frac{1}{8\pi G}\int_{\Sigma\cap {\cal N}}\alpha(\delta_1,\delta_2)\ =\ \frac{1}{4\pi G} \delta_{[1}a \delta_{2]}(\kappa) v .
\end{equation}

\subsection{The $U(1)$ Chern-Simons formulation}\label{sectU1}
Here we will discuss in more detail the gauge fixing introduced in the preceding section. Later, a horizon section area $a$ will be fixed to be constant in the relevant part of the phase space, but we will first allow WIH of all the possible areas and coframes constrained only by the gauge fixing conditions GF1-GF3 listed above.

For the assumed class of frames the right hand side of (\ref{Omega}) is
\begin{equation}\label{Omega3}
2d\left(\delta_{[1}(e^1\wedge e^2)^{({\cal N})} \delta_{2]}(\kappa) v\right)\ -\   \frac{2}{\gamma}\left(\delta_{[1}(e^1\wedge e^2)\wedge \delta_{2]}\Gamma_{12}\right)^{({\cal N})}.
\end{equation}

To better understand the meaning of the connection component  $\Gamma_{12}$ pulled back to ${\cal N}$, consider first the case of a frame  ${\bar{e}}_I$ such that
$${\cal L}_\ell\bar{e}^{\,1}\ =\ {\cal L}_\ell\bar{e}^{\,2}\ =\ 0.$$
Then, the corresponding connection 1-form component  satisfies
$$\bar{\Gamma}_{12a}\ell^a\ =\ 0,\ \ {\cal L}_\ell (\bar{\Gamma}_{12})^{({\cal N})}=0$$
and hence  $(\bar{\Gamma}_{12})^{({\cal N})}$ can be identified with its pullback onto a slice $S_v$, that is,  the Levi-Civita connection 1-form of the 2-metric tensor $g^{(S)}$ defined on  any cross-section\footnote{Remember that any two cross-sections $S_v$ and $S_{v'}$ are naturally isometric to each other.} $S=S_v$ of ${\cal N}$
by the orthonormal co-frame $((\bar{e}^{\,1})^{(S)},(\bar{e}^{\,2})^{(S)})$.
The general form of the coframes which satisfy the conditions GF1-GF3 of Section \ref{setup} is
\begin{equation} e^A=r^A_B\bar{e}^{\,B},\ \ \ A,B=1,2,\ \ \ e^3=\bar{e}^{\,3},\ \ \ e^4=\bar{e}^{\,4}, \end{equation}
where $r^A_B=r^A_B(x)$ is a $SO(2)$ matrix depending, generically, on the point $x\in{\cal N}$.  Due to the transformation law for gauge connections, we have
$$(\Gamma^1{}_2)^{({\cal N})}\ =\ \left(\bar{\Gamma}^1{}_2\right)^{({\cal N})}\ + dh$$
where
$$dh\ =\ r^1_Bd(r^{-1}){}^B_1.$$
Therefore
\begin{equation} d\left(\Gamma^1{}_2\right)^{({\cal N})}\ =\ d\left(\bar{\Gamma}^1{}_2\right)^{({\cal N})}\ =\ \frac{1}{2}R\left(e^1\wedge e^2\right)^{({\cal N})} \end{equation}
where $R$ is the Riemann scalar curvature of the 2-geometry of $S$.

It turns out that the second term in (\ref{Omega3}) can be written in the form corresponding to a $U(1)$ Chern-Simons theory after a suitable choice of variables. Specifically, the suitable variable is a $SO(2)$ connection $A$ on ${\cal N}$ with constant curvature equal to the average curvature of $\left(\Gamma^1{}_2\right)^{({\cal N})}$. In other words, such that
\begin{equation}\label{dA} dA\ =\ \frac{1}{2}\langle R\rangle \left(e^1\wedge e^2\right)^{({\cal N})} \end{equation}
where
 \begin{equation} \langle R\rangle\ =\ \frac{1}{a}\int_{S_{v}} R e^1\wedge e^2. \end{equation}
That is,
$$ A\ =\ \left(\Gamma_{12}\right)^{({\cal N})}\ + w $$
 where $w$ is a globally\footnote{As opposed to $\left(\Gamma^1{}_2\right)^{({\cal N})}$ which is defined only  locally, i.e. in the chart where the local frame $(e^1,e^2)$ is defined.} defined 1-form on ${\cal N}$ satisfying
$$ dw\ =\ \frac{1}{2}(R-\langle R\rangle) \left(e^1\wedge e^2\right)^{({\cal N})},\ \ \ {\cal L}_\ell w\ =\ 0\\,,\ \ \ \ \ell^aw_a=0\,. $$
To find $w$ we must notice first that the global existence of $w^{(S)}$ on each $S$ follows from the vanishing of the integral of the right hand side of the the above equality, namely
$$  \frac{1}{2}\int_{S_v}(R-\langle R\rangle) e^1\wedge e^2\ =\ 0.  $$
Next, $w$ is defined on ${\cal N}$ by requiring it to be orthogonal to, and Lie dragged by, $\ell$. The connection $A$ is defined on the same bundle as $\Gamma^1{}_2$, that is, on the orthonormal frame bundle. In order to agree with \cite{ABE,BE} we define
\begin{equation}\label{V} V\ :=\ -\frac{1}{2}A\ \end{equation}
which is a $U(1)$ connection on the square root of the orthonormal frame bundle over $S$.

Finally, in terms of the connection $V$ it is easy to see that
$$ \delta_{[1}(e^1\wedge e^2)^{({\cal N})}\wedge \delta_{2]}(\Gamma_{12})^{({\cal N})}\ =\ 8\delta_{[1}\left( \frac{dV}{\langle R\rangle}\right)\wedge \delta_{2]}V.  $$
The average curvature $\langle R\rangle $ can be obtained by using the topological invariant
$$\label{8pi} \int_{S_v} R e^1\wedge e^2\ =\ 8\pi$$
as
$$\langle R\rangle\ =\ \frac{8\pi}{a}. $$
Therefore, for variations $\delta_1$ and $\delta_2$ such that
$$   \delta_1 a\ =\ \delta_2 a\ =\ 0\,, $$
we have
\begin{equation}\label{CS} \delta_{[1}(e^1\wedge e^2)^{({\cal N})}\wedge \delta_{2]}(\Gamma_{12})^{({\cal N})}\ =\ d\left(\frac{a}{4\pi}\delta_{1}V\wedge \delta_{2}V\right).  \end{equation}

\begin{figure}[htbp]
\includegraphics[width=15.2cm]{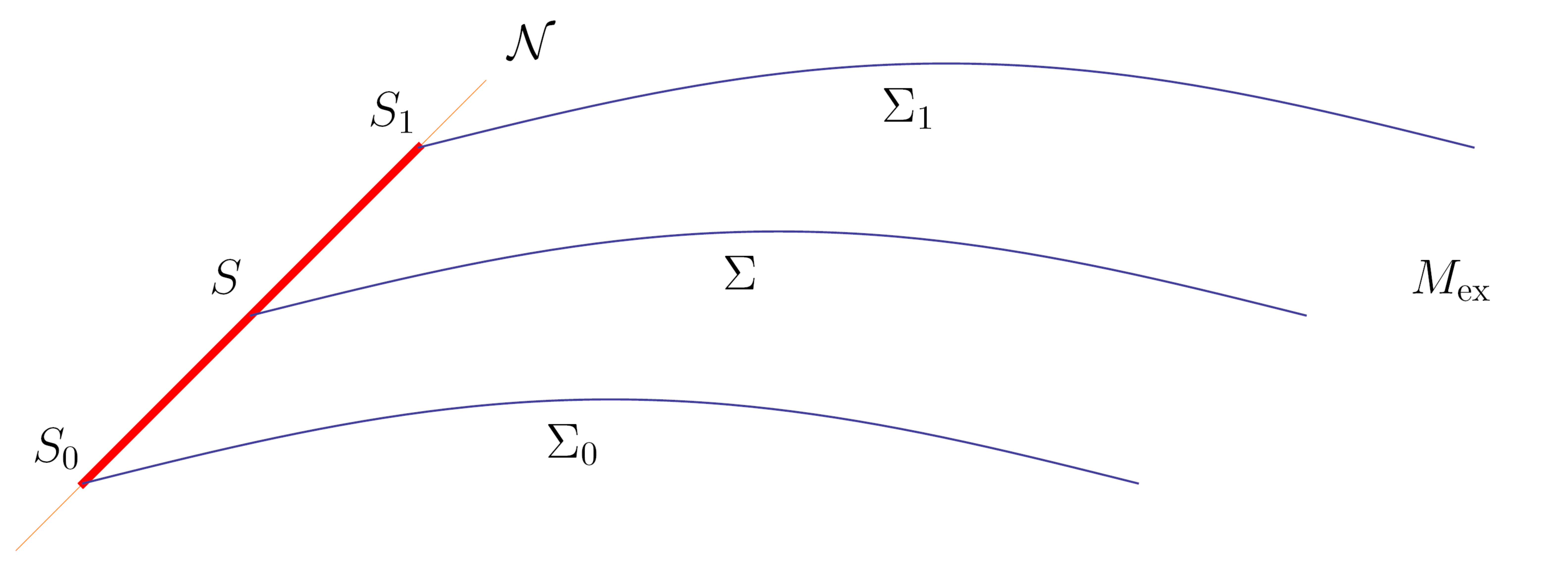}
\caption{Schematic picture of the region $M_{\mathrm{ex}}$.} \label{F0}
\end{figure}

The final formulation of the theory is as follows. We consider the space  of spacetimes each of which is the exterior of a weakly isolated horizon. These spacetimes satisfy the Einstein vacuum equations. The areas of all the horizons are equal to an arbitrarily fixed number $a$. The spacetime geometries are defined in a region $M_{\rm ex}$ of a 4-dimensional
manifold $M$ contained between 3-surfaces ${\cal N}$, $\Sigma_0$ and $\Sigma_1$  (see Figure \ref{F0}).

We will build now a coordinate system $(x^1,x^2,v)$ on ${\cal N}$ by pulling back coordinates on $S^2\times (v_0,v_1)$. In order to do this let us fix a diffeomorphism
\begin{equation}\label{diffeo}
{\cal N}\ \rightarrow\ S^2\times (v_0,v_1)\ ,
\end{equation}
where $S^2$ is a 2-sphere, on which we have selected a chart with coordinates $(x^1,x^2)$ and $v\in(v_0,v_1)$.
The interpretation of $v$ is clear now, $M_{ex}$ is foliated by 3-manifolds $\Sigma$ whose intersections with ${\cal N}$, given by the condition $v = $ constant, are diffeomorphic to a 2-sphere. We restrict ourselves to metric tensors $g$ on $M$ such that $({\cal N}, [{\partial_v}])$ is a weakly isolated horizon in which the lines $x^{1,2}=$constant are null geodesics affinely parametrized by $v$ (the parametrization of $[v_0,v_1]$) and the surfaces $\Sigma$ are spacelike. Each of the allowed metric tensors $g$ is represented by coframes $e^I$ defined on open domains covering $M$ such that
$$g=e^1\otimes e^1\ +\ e^2\otimes e^2\ - e^3\otimes e^4\ -\ e^4\otimes e^3.$$
Furthermore, each coframe satisfies at ${\cal N}$ the conditions introduced in Section \ref{setup}, that is,
$$e_4\ =\ \partial_v,$$
$e_1$ and $e_2$ are tangent to the 2-surfaces $S_v$ defined by the condition $v=$const.

According to (\ref{integralN}), (\ref{Omega}), (\ref{Omega3}) and (\ref{CS}), the symplectic form in the space of geometries is now
\begin{eqnarray*}
\Omega(\delta_1,\delta_2)=\Omega_{\rm Hor}(\delta_1,\delta_2)\ +\ \Omega_{\rm Bul}(\delta_1,\delta_2)
\end{eqnarray*}
where
\begin{eqnarray*}
\Omega_{\rm Hor}(\delta_1,\delta_2)&:=& \frac{a}{2(2\pi)^2 G\gamma}\int_{S}\delta_{1}V\wedge \delta_{2}V\,, \nonumber\\
\Omega_{\rm Bul}(\delta_1,\delta_2)&:=&
\frac{1}{8\pi G}\int_{\Sigma}\left(\frac{1}{2}\epsilon_{IJKL}\delta_{[1}(e^I\wedge e^J)\wedge \delta_{2]}\Gamma^{KL}\ -\ \frac{1}{\gamma}\delta_{[1}(e^I\wedge e^J)\wedge \delta_{2]}\Gamma_{IJ}\right)\,. \nonumber
\end{eqnarray*}
We can simplify the expression of the symplectic form in the bulk (that is in $\Sigma$) by introducing a condition restricting the Lorentz transformations of the frame. This is done by assuming that $e_1,e_2$ and
$$e_I r^I\ =\ \frac{1}{\sqrt{2}}(e_4-e_3)$$ are tangent to $\Sigma$, that is fixing the frame normal to $\Sigma$ to be
$$e_I n^I\ =\ \frac{1}{\sqrt{2}}(e_4+e_3). $$
The bulk part of $\Omega$ takes then the following form
\begin{equation} \frac{1}{8\pi G}\int_{\Sigma}\left(\frac{1}{2}\epsilon_{IJKL}\delta_{[1}(e^I\wedge e^J)\wedge \delta_{2]}\Gamma^{KL}\ -\ \frac{1}{\gamma}\delta_{[1}(e^I\wedge e^J)\wedge \delta_{2]}\Gamma_{IJ}\right)\ =\ 2\int_{\Sigma}\partial_{[1} P^a_i\partial_{2]} A^i_a  \end{equation}
where we have used  real Ashtekar variables which emerge from this symplectic form quite naturally because
\begin{align} P^a_i\ &=\ \frac{1}{16\pi G\gamma}\epsilon^{abc}\epsilon_{IJKi}n^I(e^J)^{(\Sigma)}_b(e^K)^{(\Sigma)}_c\,,\\
A^i_a\ &=\   (\Gamma^{KL})^{(\Sigma)}_a(\frac{1}{2}\epsilon^i{}_{JKL}\ +\ \gamma\delta^i_J)n^J \,,
\end{align}
where $P_i^a$ stands for $P_1^a$, $P_2^a$, and $P_I^ar^I$ and similarly for $A^i_a$.

In this approach, the fields $V$ on $S$ and $P$, $A$ on $\Sigma$ are subject to the {\it horizon constraint} \cite{Ashtekar:1999wa}
\begin{equation} \label{theconstraint1} dV^{(S)}\ +\  \frac{(4\pi)^2G\gamma}{a}\epsilon_{abc}P^a_Lr^L(dx^b\wedge dx^c)^{(S)}\ =\ 0.
\end{equation}
The constraint (\ref{theconstraint1}) has a clear geometric meaning. It generates the gauge transformations defined by the rotations at each tangent space to $S$
\begin{equation} e'_1\ =\ \cos 2\Lambda\, e_1\ +\ \sin 2\Lambda\, e_2,\ \ e'_2\ = -\sin 2\Lambda\, e_1\ +\ \cos 2\Lambda\, e_2, \
e'_Ir^I \ =\ e_Ir^I  \end{equation}
where $\Lambda:S\rightarrow \mathbb{R}$ is an arbitrary differentiable function, namely
\begin{align} V'\ &=\ V + d\Lambda,\ \ \ \  \ \ \ \ \ \ \ \ \ \ \ \ A'^Ir_I\ =\ A^Ir_I\ +\ 2d\Lambda,\nonumber\\
A'^1\ &=\ \cos 2\Lambda\, A^1\ -\ \sin 2\Lambda\, A^2,\ \ \ \ A'_2\ = \sin 2\Lambda A_1\ +\ \cos 2\Lambda\, A_2, \nonumber\\
 P'_1\ &=\ \cos 2\Lambda\, P_1\ +\ \sin 2\Lambda\, P_2,\ \ P'_2\ = -\sin 2\Lambda\, P_1\ +\ \cos 2\Lambda\, P_2, \ \
P'_Ir^I \ =\ P_Ir^I.
    \end{align}
This constraint will be used below in the integrated form, namely, for every non-self intersecting loop $\partial s$ in $S$ which has the interior $s\subset S$, the integrated version of the horizon constraint reads
\begin{equation}\label{theconstraint2} \frac{a}{2(4\pi)^2G\gamma}\int_{\partial s} V\ +  P_{s,r}\ =\ 0\end{equation}
where
\begin{equation}\label{flux} P_{s,r}\ :=\ \int_{s}P^a_Lr^L \frac{1}{2}\epsilon_{abc}dx^b\wedge dx^c
\end{equation}
is the flux  of the vector density $P^a_I r^I$ along the 2-surface $s$.

\subsection{The $SU(2)$ Chern-Simons formulation}
In addition to the $U(1)$ framework described before there is also a recent $SU(2)$ approach for vacuum gravity which admits as symmetries all the rotations on the horizon \cite{Engle:2009vc,Engle:2010kt,Engle:2011vf}. In this approach one assumes the {\it spherical symmetry} of the spacetime geometry on the \emph{isolated horizon} and consider only horizons with a fixed area. An isolated horizon (IH) is a WIH satisfying the additional condition that
$$
[\mathcal{L}_\ell,D]=0\,,
$$
where  $D$ is the intrinsic derivative operator defined in $\cal N$ by $X^aD_aY^b := X^a\nabla_aY^b$, where $X$, $Y$ are vector fields tangent to $\cal N$ and $\nabla$ the Levi-Civita connection associated with $g$. Notice that $D$ is well defined because it preserves the tangent bundle of ${\cal N}$ as a consequence of $\mathcal{L}_\ell g^{({\cal N})}=0$.

In this case the frame (gauge) fixing conditions are somewhat relaxed because neither $e_4$ is assumed to be $\ell$ nor $e_1$ and $e_2$ are tangent to $S$. However, the time gauge is still fixed, that is, $n^Ie_I$ defined in the previous subsection  is still assumed to be orthogonal to $\Sigma$.  Upon those assumptions the horizon part of the symplectic form becomes
\begin{equation}\Omega_{\rm Hor}(\delta_1,\delta_2)\ =\ \frac{a}{8\pi^2(1-\gamma^2)\gamma}\int_S\delta_1A^i\wedge \delta_2A^i\,,\end{equation}
which corresponds to that of a $SU(2)$ Chern-Simons theory. Finally the form of the \textit{horizon constraint} becomes now
\begin{equation}\label{theconstraint3} \Big(\frac{1}{2}P_i^a\epsilon_{abc}dx^a\wedge dx^b\ +\ \frac{a}{8\pi^2(1-\gamma^2)\gamma}F_{i}\Big)^{(S)}\ =\ 0\,,\end{equation}
where $P_i$ and $A^i$ are the same as above, and $$F^i\ =\ dA^i+\frac{1}{2}\epsilon^i_{jk}A^j\wedge A^k.$$
The constraint (\ref{theconstraint3}) generates $SU(2)$ gauge transformations on $S$ and coincides, modulo some factors involving the Immirzi parameter, with the constraint given in \cite{Ashtekar:1997yu} before performing the customary partial gauge fixing leading to the $U(1)$ model. Notice, however \cite{Alej}, that the remaining two conditions obtained by projecting (\ref{theconstraint3}) with respect to two independent internal directions orthogonal to $r^I$, together with the gauge fixing conditions GF1-GF3 given in Section \ref{setup}, should be solved. This must be done classically (i.e. before quantizing) though this was not the path followed in the original papers on the subject where, instead, the additional conditions were implemented \textit{weakly} in the quantized version. This, together with the fact that the $U(1)$ model does not use spherical symmetry, are some of the reasons that explain the discrepancy of both approaches.

\section{Quantum isolated horizons}

The Hamiltonian analysis of general relativity in the presence of isolated horizon inner boundaries leads to the introduction of a Hilbert space built as the tensor product of a bulk Hilbert space and a horizon Hilbert space,  $\mathcal{H}^k_{\rm Kin}=\mathcal{H}^k_{\rm Hor}\otimes\mathcal{H}_{\rm Bul}$.  Here the bulk degrees of freedom are associated with the Cauchy surface $\Sigma$ and the horizon degrees of freedom  with the intersection of $\Sigma$ with the isolated horizon $S=\Sigma\cap\mathcal{N}$. The bulk Hilbert space admits a basis spanned by spin networks that are allowed to intersect the horizon $S$. These intersections are known as \textit{punctures} and carry quantum numbers $j_I$, labeling $SU(2)$ irreducible representations  associated with the intersecting edges. They also carry quantum numbers $m_I$ defined by projecting the spin vectors with respect to any space-like vector field on the horizon (that can be defined with the help of extra privileged structures when available). In any case the horizon Hilbert space $\mathcal{H}^k_{\rm Hor}$ corresponds to a Chern-Simons (CS) theory with level $k\in \mathbb{N}$. As mentioned above, this may have either $U(1)$ or $SU(2)$ as gauge group depending on the details of the treatment. The difference between them is the different role played by symmetry requirements and the treatment of the \textit{quantum boundary conditions}. These are the quantum counterparts of equations (\ref{theconstraint2}) and (\ref{theconstraint3}), respectively. They generate the gauge transformations on the horizon. The classical isolated horizon boundary conditions imply that the connection is reducible on the horizon. This means that there exist internal vectors on the sphere such that their covariant derivatives are zero. As explained in section \ref{sectU1}, a particular choice $r^I$ for this vector can be interpreted as a partial gauge fixing condition for the $SU(2)$ connection that implements a symmetry reduction from $SU(2)$ to $U(1)$. This is the path followed in the original treatment \cite{Ashtekar:1997yu}, where the only component of the boundary conditions that is promoted to a quantum operator and implemented \textit{\`a la Dirac} is the $r^I$-projection of equation (\ref{theconstraint3})  (see \cite{Basu:2009cw} for a discussion on this issue). On the other hand, in the $SU(2)$ treatment the full set of conditions are promoted to quantum operators and enforced on the physical states without any gauge fixing.

In the $U(1)$ case \cite{Ashtekar:1997yu,Ashtekar:2000eq} the horizon Hilbert space ${\cal H}^k_{\rm Hor}$ depends on the level $k\in \mathbb{N}$ of the quantum CS-theory. This fixes the prequantized value of the area of the isolated horizon to the value $a_k=4\pi\gamma\ell^2_P k$. The Hilbert space is spanned by $U(1)$-CS basis states $|(c_1,\ldots,c_N)\rangle^k_{\rm Hor}$ defined on the punctured sphere. They are labeled by ordered sequences of non-zero congruence classes of integers modulo $k$. Each of the labels $c_I$ is an integer number in the set $\{1,2,\ldots,k-1\}$ and labels the quantized deficit angle $4\pi c_I/k$ of the $I$-th puncture. The spherical topology of the horizon $S$ imposes an additional restriction on the curvature that translates into the following condition
$$\sum_I c_I=0 \quad  (\mathrm{mod} \,\, k),$$
for the  labels in a given sequence $(c_1,\ldots,c_N)$.  The  CS-labels $(c_I)$ are related, via the quantized isolated horizon boundary condition (\ref{theconstraint2}), to the \textit{quantum geometric} labels $m_I$ corresponding to the $j_I$ representation of the edge piercing the horizon at the corresponding puncture. The condition that they must satisfy is
$$c_I=-2m_I \quad (\mathrm{mod}\,\,k)\ ,$$
where $m_I\in\left\{-j_I, -j_I+1,\ldots,j_I\right\}$. This restriction on the form of the basis states $|(c_I)\rangle_{\rm Hor}^k\otimes |(j_I,m_I),\cdots\rangle_{\rm Bul}$ of $\mathcal{H}^k_{\rm Kin}$ is the quantum counterpart of the isolated horizon boundary condition.

In the $SU(2)$ proposal \cite{Engle:2009vc,Engle:2010kt} the horizon Hilbert space is that of a $SU(2)$ Chern Simons theory with a level $k$ corresponding to the same prequantized value of the area. The quantum states are labeled now by representations $s_I$ of the quantum group $SU(2)_q$. The quantum matching conditions, (\ref{theconstraint3}) in this case, leads to an identification of these labels with the $j_I$, that is, the labels associated with the edges of the spin network piercing the horizon at the punctures.

\section{Black hole entropy}

The quantum states of the horizon belonging to $\mathcal{H}_{\rm Hor}^k$ and compatible with a given value of the area are responsible for the black hole entropy. Given a bulk state vector labeled by a spin network piercing the isolated horizon, it is possible to assign an area to the horizon as the eigenvalue of the area operator given by
\begin{eqnarray}
\label{area} a^{{\scriptscriptstyle \rm LQG}}(j_I,m_I)=8\pi\gamma\ell_P^2\sum_I{\sqrt{j_I(j_I+1)}}\,,
\end{eqnarray}
where the $j_I$ are the labels of the edges at the punctures. At this point it is important to mention that two types of areas related to the horizon. The first is the prequantized area $a_k$ that must be introduced in the quantization of the CS theory. The second is the area eigenvalue assigned to the horizon by the spin network and given by (\ref{area}). In principle it would be desirable to take both areas as equal. However, the fact that $a_k$ is not an element of the area spectrum precludes us from doing this.\footnote{We want to point out, nonetheless, the existence of other choices for the horizon area operator \textit{within} the LQG approach \cite{FernandoBarbero:2009ai}, with evenly spaced area eigenvalues, in which this problem disappears.} In practice this difficulty is sidestepped by introducing an area interval $[a_k-\delta,a_k]$ with a large enough $\delta$ ensuring the presence of a prequantized value of the area in it. Though this may seem an \textit{ad hoc} way to solve this problem, the introduction of certain intervals is, in fact, customarily used in statistical mechanics to define some relevant ensembles (the microcanonical in particular). The ultimate reason why this is acceptable relies on the fact that, in an appropriate thermodynamic limit, the value of the entropy is independent of the width of the interval. Actually, it is possible to consider intervals of the form $[a_0,a_k]$ (with $a_0$ being the lowest area eigenvalue) to simplify the computations. The consideration of the thermodynamic limit is also important for conceptual reasons related to the smoothness properties of the entropy. To be usable in standard thermodynamics the entropy must satisfy some regularity requirements, for example it must be a differentiable function of the energy (otherwise it is impossible to define the temperature). As shown in the classic paper by Griffiths \cite{Griffiths} this is guaranteed in the thermodynamic limit.

An important comment to be made at this point concerns the use of area ensembles. The standard framework of statistical mechanics is based on the use of \textit{energy} ensembles. Both the standard microcanonical and canonical ensembles are defined in terms of the energy. The statistical entropy of a gas is obtained, for example, by counting the number of energy eigenstates below a given fixed energy value $E$. The partition function, in its stead, is obtained by adding $e^{-\beta E_n}$ for all the possible energy eigenvalues (taking into account their degeneracies). In the case of black holes in LQG the role of the energy is played by the area \cite{Krasnov:1996tb}. This is somehow a necessity in the formalism as this is the only geometric/physical quantity that can be assigned in a natural way to the horizon by a given bulk spin network (see however \cite{Alej} for a more elaborate point of view on this issue). The actual definition of the black hole entropy is performed by tracing out over the bulk states to obtain a density matrix describing a maximal entropy mixture of surface states with eigenvalues in the area interval of our choice.

In order to obtain the statistical entropy in the area microcanonical ensemble the relevant combinatorial problem in the $U(1)$ framework, as explained at length in \cite{Ashtekar:2000eq}, consists on counting the sequences $(c_I)$ of non-zero elements of $\mathbb{Z}_k$ satisfying $c_1+\cdots+c_N=0$, and such that the condition  $c_I=-2m_I\, (\mathrm{mod}\, k)$ is satisfied for \textit{permissible} spin components $(m_I)$. In this context we say that such a sequence of $m_I$ labels is permissible if there exists a sequence of non-vanishing spins $(j_I)$ such that each $m_I$ is one of the spin components of $j_I$ and
\begin{eqnarray}
a_k-\delta\leq a^{{\scriptscriptstyle \rm LQG}}(j_I,m_I)=8\pi\gamma\ell_P^2\sum_I\sqrt{j_I(j_I+1)}\leq a_k.
\end{eqnarray}
In practice, the counting of  $c$-labels amounts to the determination of the dimension of the Hilbert subspace of ${\cal H}^{k}_{\rm Hor}$ describing the black hole degrees of freedom.

A final, and somewhat unexpected, advantage of introducing an area interval is the possibility of simplifying the actual computation of the entropy by restating the combinatorial counting problem in simpler terms. Specifically, by employing an interval of the form $[0,a_k]$, Domagala and Lewandowski \cite{Domagala:2004jt} proved that the black hole entropy can be obtained by the following prescription involving only the \textit{bulk} labels $m_I$:
\bigskip

\noindent{\textbf{DL-black hole entropy} \cite{Domagala:2004jt}:}
\textit{The entropy $S^{\mathrm{DL}}_{\mathrm{stat}}$ of a quantum horizon of classical area $a_k$ according to the Ashtekar-Baez-Corichi-Krasnov (ABCK) framework is
$$
S^{\mathrm{DL}}_{\mathrm{stat}}(a_k) = \log \Omega^{\rm DL}(a_k)\,,
$$
where $\Omega^{\rm DL}(a_k)$ is 1 plus the number of all the finite, arbitrarily long, sequences $(m_1,\ldots,m_N)$ of non-zero half integers, such that the following equality and inequality are satisfied:
\begin{equation}
\sum_{I=1}^N m_I=0\,, \label{cond1}
\end{equation}
\begin{equation}
\sum_{I=1}^N\sqrt{|m_I|(|m_I|+1)}\leq \frac{a_k}{8\pi\gamma\ell_P^2}\,.\label{cond2}
\end{equation}
The extra term 1 above comes from the trivial sequence.}

\bigskip

To show that this simplification is possible one has to build a bijection between the sequences $(c_I)$ satisfying the conditions made explicit above, and the sequences $(m_I)$ that appear in the preceding prescription. By doing this we show that it is equivalent to count the sequences $(c_I)$ or the $(m_I)$. Let us suppose first that we are given a sequence $(m_I)$ satisfying (\ref{cond1}) and (\ref{cond2}). If we take now $c_I=-2m_I (\mathrm{mod}\,k)$ and $j_I=|m_I|$ the sequence $(c_I)$ obviously satisfies the conditions $c_1+\cdots+c_N=0$ and $c_I=-2m_I (\mathrm{mod}\,k)$. Also, the sequence $(m_I)$ is trivially permissible (just consider $j_I=|m_I|$). Conversely, if we are given a sequence $(c_I)$ satisfying the conditions necessary for it to be counted in the computation of the entropy we can find a unique sequence $(m_I)$ satisfying (\ref{cond1}) and (\ref{cond2}). To see this we follow \cite{Domagala:2004jt}.  Suppose that we are given one of the prequantized values for the area corresponding to the CS level $k$. Now the condition $c_1+\cdots+c_N=0$ implies that
\begin{equation}
\sum_{I=1}^Nm_I=\frac{k L}{2}\,,\quad L\in\mathbb{Z}\,.
\label{summ}
\end{equation}
In order to find a permissible sequence of spin components $(m_I)$ the following chain of inequalities is useful
\begin{eqnarray}
a_k=4\pi\gamma\ell_P^2 k&\geq& 8\pi\gamma\ell_P^2\sum_{I=1}^N\sqrt{j_I(j_I+1)}\nonumber\\
&\geq& 8\pi\gamma\ell_P^2\sum_{I=1}^N\sqrt{|m_I|(|m_I|+1)}\nonumber\\
&>& 8\pi\gamma\ell_P^2\sum_{I=1}^N|m_I|>8\pi\gamma\ell_P^2\big|\sum_{I=1}^N m_I\big|\nonumber\\
&=& 4\pi\gamma\ell_P^2k |L|\,,
\end{eqnarray}
where we have made use of (\ref{summ}) in the last step. As a consequence we see that $L$ must vanish and, hence, the sought for sequence $(m_I)$ must satisfy the condition (\ref{cond1}), known in the literature as the \textit{projection constraint}. Another inequality that we can read off from the previous chain is
$$
\sum_{I=1}^N|m_I|<\frac{k}{2}\,,
$$
which implies that $|m_I|<k/2$ for $I=1,\ldots,N$. This restriction on the possible values for each $m_I$ is such that there is only a single choice for $m_I$ that satisfies the condition $c_I=-2m_I\,(\mathrm{mod}\,k)$ for the given $c_I$. We conclude then that there is a unique permissible sequence $(m_I)$ associated with the given $(c_I)$ and also that it must satisfy the projection constraint.

\noindent Notice that the entropy $S^{\mathrm{DL}}_{\rm stat}(a)$ is defined \textit{only} for area the prequantized area values $a_k$. However, following the proposal by \cite{Corichi:2006bs,Corichi:2006wn} we will extend the definition to arbitrary values of $a\in[0,\infty)$ by just requiring that
$$
\sum_{I=1}^N\sqrt{|m_I|(|m_I|+1)}\leq \frac{a}{8\pi\gamma\ell_P^2}\,.
$$
This extension is partially justified by the fact that something similar is done in the standard treatments of the microcanonical ensemble and also because the detailed form of the area spectrum of loop quantum gravity is one of the main predictions of the formalism. In the following we will write areas in units of $4\pi\gamma\ell_P^2$ unless stated otherwise. It is important to stress at this point that the foregoing rephrasing of the combinatorial problem that must be solved to compute the entropy does not change the fact that what we really want to count are some Chern-Simons states on the horizon. The use of the $m_I$ labels is a (very) useful simplifying device but nothing more.

A reasoning along the lines presented above leads to the statement of the combinatorial problem in the $SU(2)$ case. Now the entropy is obtained by counting states in the $SU(2)$ Chern-Simons phase space. The quantum boundary condition and the introduction of an area interval lead to a combinatorial problem in which one has to count sequences of spin labels $(j_I)$, associated with the spin network edges piercing the horizon with a degeneracy factor related to the dimension of the invariant subspace of $\mathrm{Inv}(\otimes_I[j_I])$. Specifically:

\bigskip

\noindent{\textbf{ENP-black hole entropy}.}
\textit{The entropy $S_{\mathrm{stat}}^{\rm ENP}(a_k)$ of a quantum horizon of the classical area $a_k=4\pi\gamma\ell^2_P k$ (when $\gamma\leq\sqrt{3}$) is defined as
$$S_{\mathrm{stat}}^{\rm ENP}(a_k) = \log \Omega^{\rm ENP}(a_k)\,,$$
where $ \Omega^{\rm ENP}(a_k)$ is 1 plus the number of all the finite, arbitrarily long, sequences $(j_1,\ldots,j_N)$ of non-zero half integers $j_I$ satisfying
$$8\pi\gamma\ell^2_P\sum_{I=1}^N\sqrt{j_I(j_I+1)}\leq a_k$$
and counted with a multiplicity given by the dimension of the invariant subspace $\mathrm{Inv}(\otimes_I[j_I])$.}

A nice feature of this approach is the fact that one can ultimately think of the spin network state representing a quantum black as one in which all the edges that pierce the horizon are bundled together in a single vertex labeled with an intertwiner belonging to the invariant subspace of $\otimes_I[j_I]$; (see \cite{Livine:2005mw} for more details). Moreover, it is important to point out that the most relevant part of the counting problem in this setting essentially coincides with the heuristic proposal by Ghosh and Mitra  \cite{Ghosh:2006ph}.

\bigskip

\subsection{Combinatorics: precision countings}

In this section we describe step by step the resolution of the combinatorial problems that must be solved to compute the entropy according to the $U(1)$ prescription (a simple modification of the procedure can be used to solve the problem for the $SU(2)$ case and other models \cite{Ghosh:2006ph,Bianchi:2010qd}).  The steps that must be followed are \cite{Agullo:2010zz}:

\bigskip

\noindent \textbf{Step 1.} Fix a given value of the area $a$ and find the possible choices of half integers $|m_I|\neq 0$ satisfying
    $$
    \sum_{I=1}^N\sqrt{|m_I|(|m_I|+1)}= \frac{a}{2}.
    $$
    Notice that we are considering the possible choices of $|m_I|$ as the elements of a multiset (and hence there is no ordering of the labels). In other words, at this stage we only find out how many times each spin component appears.

\bigskip

\noindent \textbf{Step 2.} Count the different ways to reorder the previous multisets.

\bigskip

\noindent \textbf{Step 3.} Count the different ways of introducing signs in the sequences $(|m_I|)$ in the previous step in such a way that $\sum_I m_I=0$.

\bigskip

\noindent \textbf{Step 4.} Repeat this procedure for each area eigenvalue smaller than $a$ and finally add the number of sequences obtained in each case.

\bigskip

The solution to the combinatorial problem described above can be encoded with the help of a \textit{generating function}. The coefficients of its expansion give the number of configurations that must be counted to compute the entropy. The derivation of the generating functions follows the steps given above. In fact, by following them it is not difficult to find out their final form. To see how this is done for the different models considered in the literature see \cite{Agullo:2010zz,G.:2008mj,Agullo:2009eq}.

The generating functions for the $U(1)$ and $SU(2)$ combinatorial problems are
\begin{eqnarray*}
G^{\rm DL}(z,x_1,x_2,\ldots)&=&\left(1-\sum_{i=1}^\infty\sum_{\alpha=1}^\infty (z^{k^i_\alpha}+z^{-k^i_\alpha}) x_i^{y^i_\alpha}\right)^{-1}\,,\\
G^{\rm ENP}(z,x_1,x_2,\dots)&=&-\frac{(z-z^{-1})^2}{2}\left(\displaystyle 1-\sum_{i=1}^\infty\sum_{\alpha=1}^\infty \Big( \frac{z^{k^i_\alpha+1}-z^{-k^i_\alpha-1}}{z-z^{-1}}\Big) x_i^{y^i_\alpha}\right)^{-1}\,.
\end{eqnarray*}
These generating functions depend, in principle, on a variable $z$ (introduced to take into account the projection constraint or the dimension of the invariant subspace in the $SU(2)$ case) and an infinite number of variables $x_i$ associated with some square-free numbers $p_i$ that appear when the eigenvalues of the area spectrum are written in the form $\sum_j q_j\sqrt{p_j}$ with $q_j\in\mathbb{N}$.  The infinite sequence of pairs of positive integers $(k^i_\alpha,y^i_\alpha)$ are the solutions to the Pell equations (one for each squarefree number $p_i$)
$$
(k+1)^2-p_i y^2=1\,.
$$
For finite areas we only have a finite number of squarefrees $p_j$ and we only need to consider the finite set of variables associated with them. For example \cite{Agullo:2010zz}, if we restrict ourselves to areas $a<18$, the only variables that we need to write explicitly in the generating functions are $x_1$, $x_2$, $x_3$, $x_4$, $x_5$, $x_7$, $x_9$, $x_{10}$, $x_{18}$, $x_{22}$, $x_{27}$, $x_{88}$, $x_{119}$, $x_{156}$, $x_{198}$ and in the $U(1)$ case we can write:
\begin{eqnarray*}
& & \hspace{-3mm}G^{\rm DL}(z,x_1,x_2,x_3,x_4,x_5,x_7,x_9,x_{10},x_{18},x_{22},x_{27},x_{88},x_{119},x_{156},x_{198})\\
& & \hspace{-3mm}\begin{array}{lllll}
=\big(1 & - (z^2 + z^{-2}) x_1^2 & - (z^{16} + z^{-16}) x_1^{12} &-(z + z^{-1}) x_2&  - (z^{6} + z^{-6}) x_2^4 \\
       &- (z^8 + z^{-8}) x_3^4 & - (z^4 + z^{-4}) x_4^2&  - (z^7 + z^{-7}) x_5^3 &- (z^9 + z^{-9}) x_7^3 \\
       &- (z^{14} + z^{-14}) x_9^4 &- (z^3 + z^{-3}) x_{10} & - (z^{10} + z^{-10}) x_{18}^2 &- (z^5 +
       z^{-5}) x_{22} \\
       &- (z^{12} + z^{-12}) x_{27}^2  & \!\!- (z^{11} + z^{-11}) x_{88}& \!\!- (z^{13} + z^{-13}) x_{119} & \!\!- (z^{15} +
       z^{-15}) x_{156} - (z^{17} + z^{-17}) x_{198}\big)^{-1}\,.
       \end{array}
\end{eqnarray*}

The coefficients $[z^0][x_1^{q_1}\cdots x_{198}^{q_{198}}]G^{\rm DL}$ tell us the values of the number of configurations obtained after following the first three steps in the previous procedure for areas of the form $a=q_1\sqrt{2}+\cdots+q_{198}\sqrt{323}$ (where some of the integer coefficients $q_i$ are allowed to be zero).  For example, for  $a=4\sqrt{2}+\sqrt{3}+\sqrt{15}$ we get
$[z^0][x_1^4x_2 x_{10}]G^{\rm DL}=24\,$.

\begin{figure}[htbp]
\includegraphics[width=15.2cm]{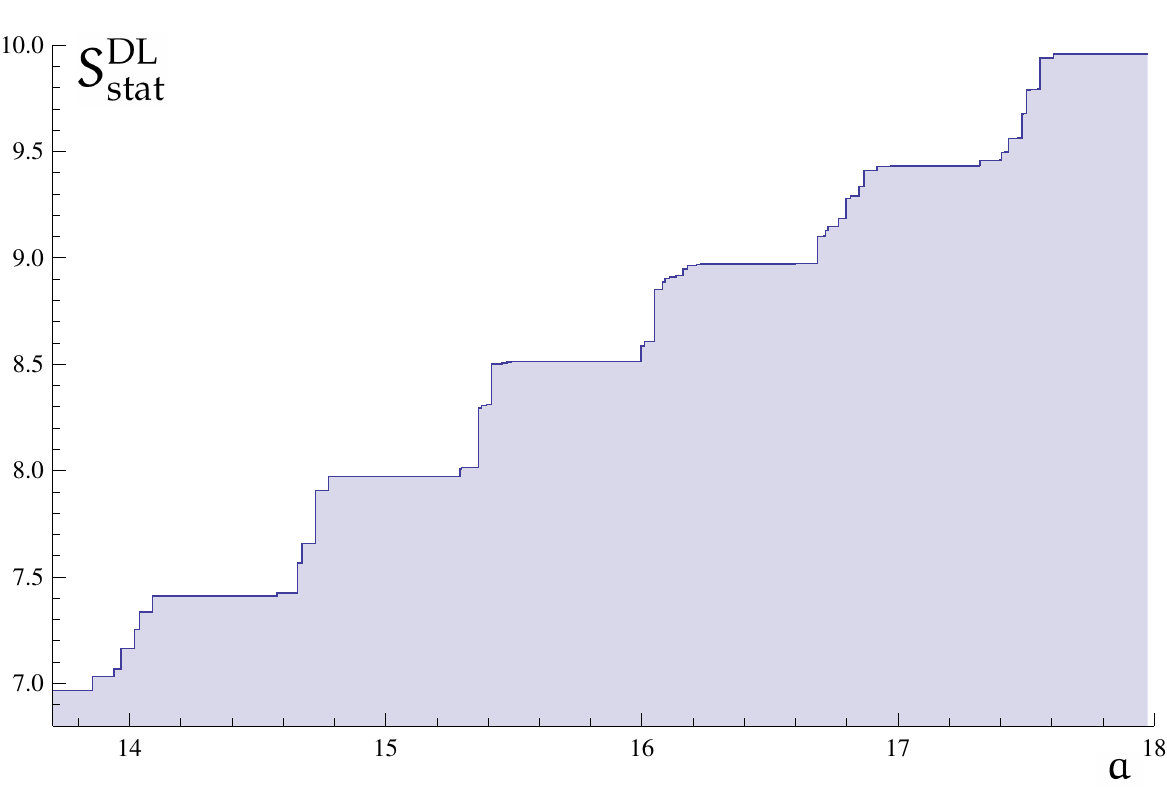}
\caption{Plot of the black hole statistical entropy $S^{\mathrm{DL}}_{\mathrm{stat}}$ in the $U(1)$ case for a range of area values between $a=14$ and $a=18$ (in units of $4\pi\gamma\ell^2_P$, where $\gamma$ is the Immirzi parameter and $\ell_P$ the Planck length). Notice the staircase structure.} \label{Fig1}
\end{figure}

These results can be directly used to plot the exact values of the entropy as a function of the area and study the detailed step structure found by the authors of \cite{Corichi:2006bs}. An example can be seen in Figure \ref{Fig1}.

The final step in the preceding list consists in adding up the numbers obtained in steps 1-3 for all the possible area eigenvalues. The most practical way to do this is by using Laplace transforms. The reason why this is so \cite{G.:2008mj} is because staircase functions such as the statistical entropy $S^{\mathrm{DL}}_{\mathrm{stat}}$ can be written as sums of Heaviside step functions of the form $\theta(a-a_0)$ (with jumps at the points $a=a_0$) with coefficients appropriately chosen to describe the height of the steps. As the location and amplitude of the steps is encoded in the generating functions given before and the Laplace transform of a step function is simply
$$
\mathcal{L}(\theta(a-a_0);s)=\frac{e^{-a_0 s}}{s}
$$
it is straightforward to find an integral representation for the entropy as an inverse Laplace-Fourier transform (the Fourier part is related to the implementation of the projection constraint in the $U(1)$ case or the counting associated with the dimension of the invariant subspace in the $SU(2)$ one). The relevant expressions for the entropy are
\begin{eqnarray}
\exp S^{\rm DL}_{\rm stat}(a)&=&\frac{1}{(2\pi)^2 i}\int_{0}^{2\pi} \int_{x_0-i\infty}^{x_0+i\infty}
\hspace{-4mm}s^{-1}\Big(\displaystyle 1-2\sum_{k=1}^\infty e^{-s\sqrt{k(k+2)}}\cos \omega k\Big)^{\!\!-1} \hspace{-2mm}e^{as}\,\mathrm{d}s \,\mathrm{d}\omega\,,\label{entroDL}
\\
\exp S^{\rm ENP}_{\rm stat}(a)&=&\frac{2}{(2\pi)^2 i} \int_0^{2\pi}\int_{x_0-i\infty}^{x_0+i\infty}\hspace{-4mm}s^{-1}\sin^2\omega \, \Big(\displaystyle 1-\sum_{k=1}^\infty \frac{\sin (k+1)\omega}{\sin \omega} e^{-s\sqrt{k(k+2)}}\Big)^{\!\!-1} \hspace{-2mm}e^{as}\,\mathrm{d}s \,\mathrm{d}\omega\,.\label{entroSU2}
\end{eqnarray}

\subsection{Asymptotics}

Two relevant questions regarding the statistical entropy are its asymptotic behavior as a function of the area (or, in other words, check that the Bekenstein-Hawking law is indeed satisfied) and the study of the persistence of the equally spaced staircase structure for large areas. In any case we want to mention at this point that the (necessary) consideration of the thermodynamic limit provides an important new perspective regarding the latest issue.

The asymptotic behavior for large areas was first derived by Meissner by looking at the singularity structure of the Laplace transform (\ref{entroDL}) that expresses the entropy \cite{Meissner:2004ju}. In the $SU(2)$ case a similar approach \cite{Agullo:2009eq} can be used. The final result of these analyses are the following asymptotic expansions for the entropy as a function of the area (measured in units of $4\pi\ell_P^2\gamma$)
\begin{eqnarray*}
S^{\scriptscriptstyle{\rm DL}}_{\mathrm{stat}}(a)&\sim& \alpha^{\scriptscriptstyle{\rm DL}}_0 a-\frac{1}{2}\log a+o(\log a)\,,
\\S^{\scriptscriptstyle{\rm ENP}}_{\mathrm{stat}}(a)&\sim& \alpha^{\scriptscriptstyle{\rm ENP}}_0 a-\frac{3}{2}\log a+o(\log a)\,.
\end{eqnarray*}
The values of $\alpha^{\scriptscriptstyle{\rm DL}}_0$ and $\alpha^{\scriptscriptstyle{\rm ENP}}_0$ are given by the position of the largest real pole of the integrands for $\omega=0$ in (\ref{entroDL}) and (\ref{entroSU2}). By choosing the value of $\gamma$ in an appropriate way it is possible to reproduce the Bekenstein-Hawking law with the right $1/4$ coefficient. It is important to realize that the coefficients of the subdominant corrections are independent of the value of $\gamma$ and \textit{differ} for the $U(1)$ and $SU(2)$ cases. These subdominant corrections are important in order to compare our results with the ones obtained in other approaches (for example in string inspired models). A good discussion of this issue appears in the interesting paper by Carlip \cite{Carlip:2000nv}.

\begin{figure}[htbp]
\includegraphics[width=15.2cm]{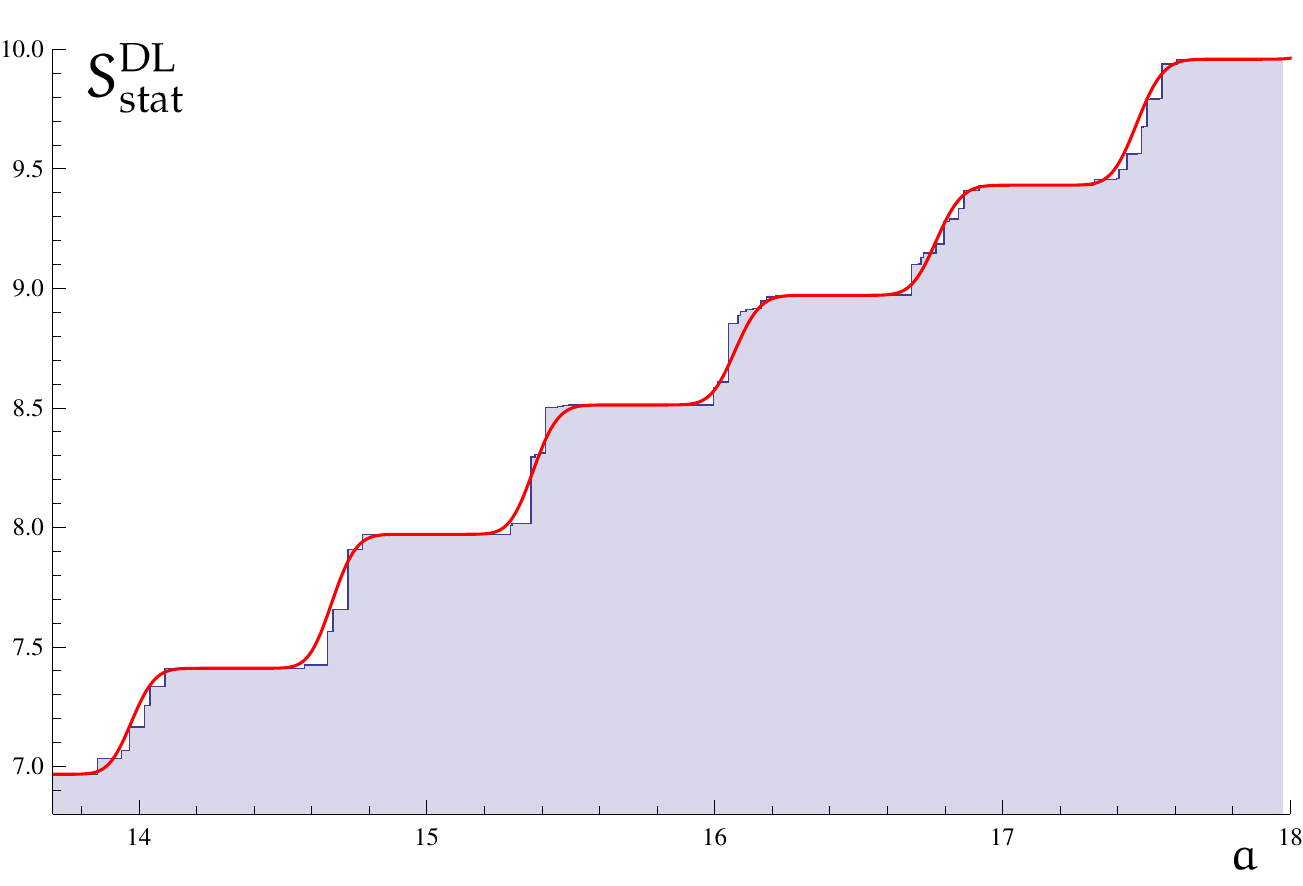}
\caption{Plot of the black hole statistical entropy $S^{\mathrm{DL}}_{\mathrm{stat}}$ in the $U(1)$ case for a range of area values between $a=14$ and $a=18$ (in units of $4\pi\gamma\ell^2_P$, where $\gamma$ is the Immirzi parameter and $\ell_P$ the Planck length). The red curve shows a smoothed approximation obtained by describing the steps with error functions as discussed in \cite{FernandoBarbero:2011kb}.} \label{Fig2}
\end{figure}

As far as the issue of the persistence of the equally spaced staircase structure is concerned, the best approach is to partition the set of black hole configurations in such a way that the individual steps are singled out. Actually there is a very efficient way to do this by using generating functions that can be directly obtained from the ones given above \cite{G.:2011zr}. In fact, it is possible to obtain generating functions for each of the individual steps. As these steps can be understood as unnormalized probability distributions it is possible to find smooth approximations for them by computing the relevant moments (the mean and the variance). The specific form of these     moments in terms of the integer label that enumerates the steps is such that the equal spacing between them is easily derived and, most importantly, the variance that measures the width of these steps grows linerly. This last fact implies that for large areas the steps are wider and wider so that the steps themselves are smoothed out in the sum (see \cite{G.:2011zr} for details). Figure \ref{Fig2} shows a comparison of the actual value of the entropy and the smoothed approximation. As can be appreciated the exact and approximate values of the entropy match almost perfectly.

\subsection{The thermodynamic limit}

The necessity to consider the thermodynamic limit can be justified in two opposite looking, but related, ways. If, for example, one studies the statistical entropy as a function of the energy in a standard thermodynamical system it is obvious that it is not a sufficiently well behaved function. In fact, the shape of the graph corresponds to a staircase function whose derivatives are either zero or are not defined at all. As important thermodynamical properties of the system are given by derivatives of the entropy (for example, the temperature) the statistical entropy ``in raw form'' cannot be used without some kind of smoothing. If instead of looking at $S_{\mathrm{stat}}$ we consider the partition function, defined as $\sum_i e^{-\beta E_i}$  (where the energy eigenvalues $E_i$ must be counted with their multiplicities) we face the opposite problem in the sense that, being an analytic function of $\beta$, it cannot be used to study such relevant issues in thermodynamics as phase transitions. The solution to both problems is the same: consider the so called thermodynamic limit in which the size of the system (as measured by some extensive parameter like the number or constituents or the volume) is taken to infinity while keeping the intensive parameters of the system fixed. Deep theorems going back to Griffiths \cite{Griffiths} show that the entropy in this limit has the desired features. It is a smooth function almost everywhere and has the right concavity properties (for reasonable interactions) to guarantee the stability of the system in question. Of course there are many possible and ad hoc ways to smooth out a staircase function. In many elementary texts this is simply done by taking all the variables that appear in concrete expressions of the entropy, including the number of particles, as continuous. However this is hardly a unique prescription and very few general statements can be derived from it.

\bigskip

\begin{figure}[htbp]
\hspace*{.2cm}
\includegraphics[width=14.5cm]{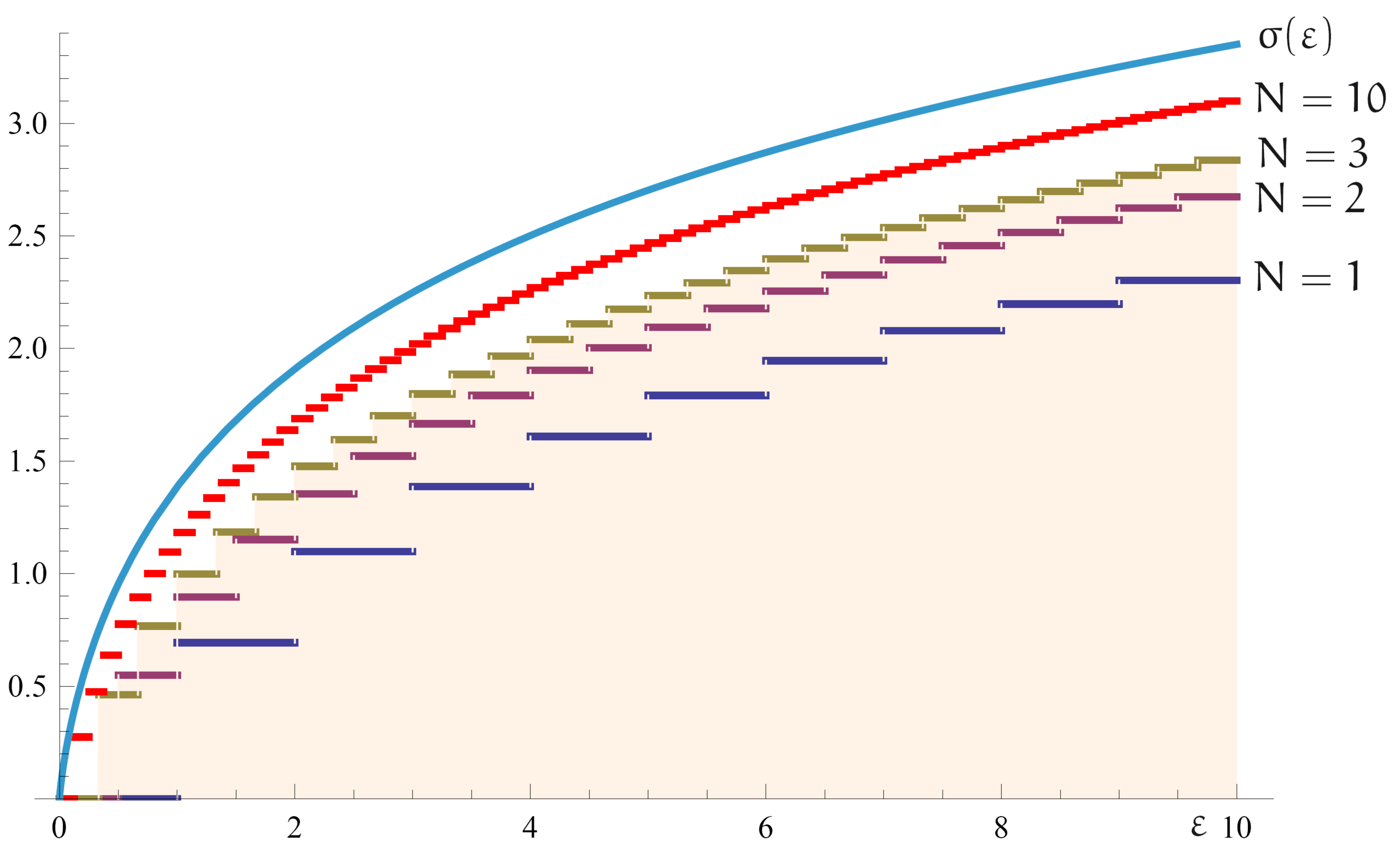}
\caption{The plot shows that, independently of the number $N$ of harmonic oscillators the statistical entropy is a staircase function whose derivatives are either zero or are not defined at all. However, the entropy in the thermodynamic limit is a smooth, concave, function of the energy with non-trivial derivatives} \label{Fig}
\end{figure}

An important difference in the behavior of the statistical and smoothed entropy can be seen in the subdominant corrections for large energies. This can be seen in illustrative examples such as the Einstein crystal that we briefly discuss now. Let us consider $N$ \textit{distinguishable}  noninteracting harmonic oscillators of angular frequency $\omega$ with (``normal ordered'') quantum Hamiltonian
$$
\hat{\mathbf{H}}(N)= \hat{H}_1+\cdots+\hat{H}_N\,,\quad\quad \hat{H}_i=\omega a_i^\dagger a_i\,.
$$
In the microcanonical ensemble, the number of eigenvectors of $\hat{\mathbf{H}}(N)$ with energy $E$ is just the number of (ordered) sequences $(n_1,\ldots,n_N)$ satisfying $\sum_{i=1}^N n_i =E/\omega$. This number can be explicitly written in terms of combinatorial numbers
$$
|\{(n_1,\ldots,n_N):n_i\in\{0\}\cup\mathbb{N}\,,\sum_{i=1}^N n_i =E/\omega\}| =\left\{\begin{array}{cl}0&\,\, \textrm{ if }\, E/\omega\not\in \mathbb{N}\\ & \\\displaystyle
\binom{N-1+\lfloor E/\omega\rfloor}{N-1}&\,\, \textrm{ if }\, E/\omega\in \mathbb{N}\end{array}
\right.
$$
Therefore, the number of microstates $\Omega(E,N)$ in an energy interval $[0,E]$ is given by
$$
S_{\mathrm{stat}}(E,N)=\log \Omega(E,N)=\log\binom{N+\lfloor E/\omega\rfloor}{N}\,.
$$
To get the thermodynamic limit, let us introduce now the energy per particle $\epsilon$ and the entropy per particle $\sigma_N(\epsilon)$,
$$
\epsilon:=\frac{E}{N},\hspace{1cm}\sigma_N(\epsilon):=\frac{S_{\mathrm{stat}}(N\epsilon,N)}{N}=
\frac{1}{N}\log\binom{N+\lfloor N\epsilon/\omega\rfloor}{N}\,,
$$
and compute the limit
$$\displaystyle
\sigma(\epsilon):=\lim_{N\rightarrow\infty}\sigma_N(\epsilon)=
    \frac{\epsilon}{\omega}\log\big(1+\frac{\omega}{\epsilon}\big)+\log\big(1+\frac{\epsilon}{\omega}\big)\,.$$
The function $\sigma(\epsilon)$ is the smoothed entropy per particle in the thermodynamic limit. If we have now $N$ particles the (extensive) smooth entropy would be
$$
S(E,N)=N\sigma(E/N)=\frac{E}{\omega}\log\big(1+\frac{N\omega}{E}\big)+N\log\big(1+\frac{E}{N\omega}\big)\,.
$$
One gets the same result with the canonical ensemble. Figure \ref{Fig} shows the statistical entropy for several values of $N$ and how they approach the entropy (per particle) in the thermodynamic limit.

\bigskip

\begin{figure}[htbp]
\includegraphics[width=15.2cm]{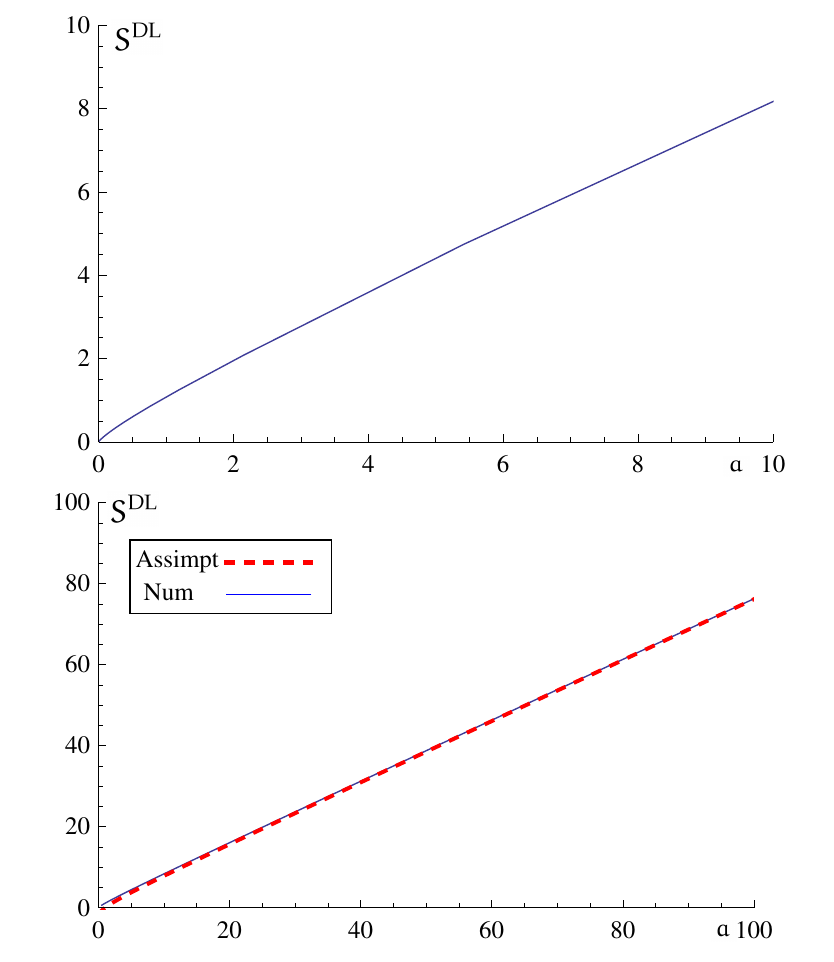}
\caption{Plot of the black hole entropy in the thermodynamical limit in the $U(1)$ case. The entropy (blue, solid curve in the second plot) is a smooth and concave curve, as can be seen in detail in the first plot. The asymptotic approximation (red, dashed curve) describes the behavior of the entropy with good accuracy.} \label{Fig3}
\end{figure}

We briefly discuss now the thermodynamic limit for black holes. In the LQG framework, where the area ensemble plays a central role, it is possible to work either with the microcanonical or the canonical ensembles. In the present case it is, in fact, simpler to rely on the canonical ensemble because the partition functions can be read off directly from the integral representations as inverse Laplace-Fourier transforms given by (\ref{entroDL}), (\ref{entroSU2}). These are
\begin{eqnarray*}
Z^{\scriptscriptstyle{\rm DL}}(\alpha)&=&\frac{1}{2\pi}\int_0^{2\pi} \frac{\mathrm{d}\omega}{1-2\sum_{k=1}^\infty e^{-\alpha\sqrt{k(k+2)}}\cos \omega k} \,,\label{ZDL}\\
Z^{\scriptscriptstyle{\rm ENP}}(\alpha)&=&\frac{1}{\pi}\int_0^{2\pi}\frac{\sin^2 \omega\,\mathrm{d}\omega}{1-\sum_{k=1}^\infty e^{-\alpha \sqrt{k(k+2)}}\sin\big( (k+1)\omega\big)/\sin\omega}\,.
\end{eqnarray*}
Once the partition function is known, the ``free energy'' --obtained by a suitable Legendre transform-- can be immediately computed \cite{G.:2011zr} and used to discuss the asymptotic behavior of the entropy as a function of the area. In particular we find that
\begin{eqnarray*}
S^{\scriptscriptstyle{\rm DL}}(a)&\sim& \alpha^{\scriptscriptstyle{\rm DL}}_0 a+\frac{1}{2}\log a+O(1)\quad a\rightarrow\infty\,.
\\
S^{\scriptscriptstyle{\rm ENP}}(a)&\sim& \alpha^{\scriptscriptstyle{\rm ENP}}_0 a+O(1)\,,\quad a\rightarrow \infty\,.
\end{eqnarray*}
In the case of the DL prescription the relevant behaviors are shown in Figure \ref{Fig3}. As it can be readily seen the entropy is a concave function of the area as expected from the general theorems mentioned above.

\section{Final comments}

The combination of the black hole model --given by the isolated horizons-- with LQG methods provides a definite setting to study quantum black holes. Although there are several proposals to carry out the quantization of these models, there is a robust formalism to study them all and extract the relevant physical information. In particular, the combinatorial problems that must be solved to compute the entropy have been thoroughly investigated and understood. The results obtained in all cases differ in some details (that may be relevant from a physical perspective) but some central predictions are robust, in particular with regard to the Bekenstein-Hawking law.

The most important issue at this point is to take into account the dynamics. This will be necessary to understand, for example, Hawking radiation (for recent work on this subject see \cite{Barrau:2011md}, other initial proposals appeared in \cite{Krasnov:1997yt}). We expect that the methods and ideas explained in the present paper will play an important role in the description of black hole evaporation. A possible setting where it seems possible to advance in this direction is the consideration of the models inspired in the Brown-Kucha\v{r} approach to the introduction of an extrinsic time variable. These have been already described within the LQG framework \cite{Giesel:2007wn,Domagala:2010bm}. Our expectation is that the general ideas and methods described here will be useful in future developments in the field and may help extract concrete predictions in the realm of black hole physics from loop quantum gravity.

\acknowledgments

We want to thank I. Agull\'o, A. Ashtekar,  E. Bianchi, E. F. Borja, A. Corichi, J. D\'{\i}az Polo, A. Perez, C. Rovelli, H. Sahlmann, and M. Varadarajan for many helpful discussions. The work has been supported by the Spanish MICINN research grants FIS2009-11893, the  Consolider-Ingenio 2010 Program CPAN (CSD2007-00042), the Polish  Ministerstwo Nauki i Szkolnictwa Wy\.zszego grants  N N202 104838 and 182/N-QGG/2008/0, and the ESF networking program QG.


\begin{thebibliography}{99}

\bibitem{Strominger}
A. Strominger and C. Vafa, \textit{Phys. Lett. B} \textbf{379} (1996) \href{http://www.sciencedirect.com/science/article/pii/0370269396003450}{99} [arXiv:hep-th/\href{http://arxiv.org/abs/hep-th/9601029}{9601029}].

\bibitem{Horowitz}
G. T. Horowitz and J. Polchinski, \textit{Phys. Rev. D} \textbf{55} (1997) \href{http://link.aps.org/doi/10.1103/PhysRevD.55.6189}{6189} [arXiv:hep-th/\href{http://arxiv.org/abs/hep-th/9612146}{9612146}].

\bibitem{Mandal}
I. Mandal and A. Sen, \textit{Class. Quant. Grav.} \textbf{27} (2010) \href{http://dx.doi.org/10.1088/0264-9381/27/21/214003}{214003} [arXiv:\href{http://arxiv.org/abs/1008.3801}{1008.3801} [hep-th]].

\bibitem{Wald:1993nt}
R. M. Wald, \textit{Phys. Rev. D} \textbf{48} (1993) \href{http://link.aps.org/doi/10.1103/PhysRevD.48.R3427}{3427} [arXiv:gr-qc/\href{http://arxiv.org/abs/gr-qc/9307038}{9307038}].

\bibitem{Iyer:1994ys}
V. Iyer and R. M. Wald, \textit{Phys. Rev. D} \textbf{50} (1994) \href{http://link.aps.org/doi/10.1103/PhysRevD.50.846}{846} [arXiv:gr-qc/\href{http://arxiv.org/abs/gr-qc/9403028}{9403028}].

\bibitem{Smolin:1995vq}
L.~Smolin, \textit{J. Math. Phys.} \textbf{36} (1995) \href{http://link.aip.org/link/doi/10.1063/1.531251}{6417}
[arXiv:gr-qc/\href{http://arxiv.org/abs/gr-qc/9505028}{9505028}].

\bibitem{Krasnov:1996tb}
K.~V.~Krasnov, \textit{Phys. Rev. D} \textbf{55} (1997) \href{http://link.aps.org/doi/10.1103/PhysRevD.55.3505}{3505} [arXiv:gr-qc/\href{http://arxiv.org/abs/gr-qc/9603025}{9603025}].

\bibitem{Krasnov:1996wc}
K.~V.~Krasnov, \textit{Gen. Rel. Grav.} \textbf{30} (1998) \href{http://www.springerlink.com/content/lq752w1170147m0u/}{53}
[arXiv:gr-qc/\href{http://arxiv.org/abs/gr-qc/9605047}{9605047}].

\bibitem{Rovelli:1996dv}
C.~Rovelli, \textit{Phys. Rev. Lett.} \textbf{77} (1996) \href{http://link.aps.org/doi/10.1103/PhysRevLett.77.3288}{3288}
[arXiv:gr-qc/\href{http://arxiv.org/abs/gr-qc/9603063}{9603063}].

\bibitem{Rovelli:1996ti}
C.~Rovelli, \textit{Helv. Phys. Acta }\textbf{69} (1996) 582
[arXiv:gr-qc/\href{http://arxiv.org/abs/gr-qc/9608032}{9608032}].

\bibitem{Ashtekar:1997yu}
A.~Ashtekar, J.~C.~Baez, A.~Corichi, and K.~V.~Krasnov, \textit{Phys. Rev. Lett.} \textbf{80} (1998) \href{http://link.aps.org/doi/10.1103/PhysRevLett.80.904}{904} [arXiv:gr-qc/\href{http://arxiv.org/abs/gr-qc/9710007}{9710007}].

\bibitem{Ashtekar:2000eq}
A.~Ashtekar, J.~C.~Baez, and K.~V.~Krasnov, \textit{Adv. Theor. Math. Phys.} \textbf{4} (2000) 1 [arXiv:gr-qc/\href{http://arxiv.org/abs/gr-qc/0005126}{0005126}].

\bibitem{Ashtekar:1999wa}
A.~Ashtekar, A.~Corichi, and K.~V.~Krasnov, \textit{Adv. Theor. Math. Phys.} \textbf{3} (2000) 419 [arXiv:gr-qc/\href{http://arxiv.org/abs/gr-qc/9905089}{9905089}].

\bibitem{Domagala:2004jt}
M.~Domagala and J.~Lewandowski, \textit{Class. Quant. Grav.} \textbf{21} (2004) \href{http://iopscience.iop.org/0264-9381/21/22/014/}{5233}
[arXiv:gr-qc/\href{http://arxiv.org/abs/gr-qc/0407051}{0407051}].

\bibitem{Meissner:2004ju}
K.~A.~Meissner, \textit{Class. Quant. Grav.} \textbf{21} (2004) \href{http://iopscience.iop.org/0264-9381/21/22/015/}{5245} [arXiv:gr-qc/\href{http://arxiv.org/abs/gr-qc/0407052}{0407052}].

\bibitem{Corichi:2006bs}
A.~Corichi, E.~F.~Borja,  and J.~Diaz-Polo, \textit{Class. Quant. Grav.} \textbf{24} (2007) \href{http://iopscience.iop.org/0264-9381/24/1/013/}{243}
[arXiv:gr-qc/\href{http://arxiv.org/abs/gr-qc/0605014}{0605014}].

\bibitem{Corichi:2006wn}
A.~Corichi, E.~F.~Borja,  and J.~Diaz-Polo, \textit{Phys. Rev. Lett.} \textbf{98} (2007) \href{http://link.aps.org/doi/10.1103/PhysRevLett.98.181301}{181301}
[arXiv:gr-qc/\href{http://arxiv.org/abs/gr-qc/0609122}{0609122}].

\bibitem{Sahlmann:2007jt} H.~Sahlmann, \textit{Class. Quant. Grav.} \textbf{25} (2008) \href{http://dx.doi.org/10.1088/0264-9381/25/5/055004}{055004} [arXiv:\href{http://arxiv.org/abs/0709.0076}{0709.0076} [gr-qc]].

\bibitem{Sahlmann:2007zp} H.~Sahlmann, \textit{Phys. Rev. D} \textbf{76} (2007) \href{http://link.aps.org/doi/10.1103/PhysRevD.76.104050}{104050} [arXiv:\href{http://arxiv.org/abs/0709.2433}{0709.2433} [gr-qc]].

\bibitem{Agullo:2008yv} I.~Agullo, J.~F.~Barbero~G.,  E.~F. Borja, J. Diaz-Polo, and E.~J.~S. Villase\~nor,
\textit{Phys. Rev. Lett.} \textbf{100} (2008) \href{http://link.aps.org/doi/10.1103/PhysRevLett.100.211301}{211301}
[arXiv:\href{http://arxiv.org/abs/0802.4077}{0802.4077} [gr-qc]].

\bibitem{BarberoG.:2008ue} J.~F.~Barbero~G. and E.~J.~S.~Villase\~nor, \textit{Phys. Rev. D} \textbf{77} (2008) \href{http://link.aps.org/doi/10.1103/PhysRevD.77.121502}{121502}
[arXiv:\href{http://arxiv.org/abs/0804.4784}{0804.4784} [gr-qc]].

\bibitem{Agullo:2010zz}
I.~Agullo, J.~F. Barbero~G.,  E.~F.~Borja, J.~Diaz-Polo, and E.~J.~S.~Villase\~nor, \textit{Phys. Rev. D} \textbf{82} (2010) \href{http://link.aps.org/doi/10.1103/PhysRevD.82.084029}{084029}
[arXiv:\href{http://arxiv.org/abs/1101.3660}{1101.3660 }[gr-qc]].


\bibitem{FernandoBarbero:2011kb}
J.~F.~Barbero~G. and E.~J.~S.~Villase\~nor, \textit{Phys. Rev. D} \textbf{83} (2011) \href{http://link.aps.org/doi/10.1103/PhysRevD.83.104013}{104013}
[arXiv:\href{http://arxiv.org/abs/1101.3662}{1101.3662} [gr-qc]].

\bibitem{G.:2011zr}
J.~F.~Barbero~G. and E.~J.~S.~Villase\~nor,  \textit{Class. Quant. Grav.} \textbf{28} (2011) \href{http://iopscience.iop.org/0264-9381/28/21/215014/}{215014}
[arXiv:\href{http://arxiv.org/abs/1106.3179}{1106.3179 }[gr-qc]].

\bibitem{Engle:2009vc}
J.~Engle, K.~Noui, and A.~Perez,
\textit{Phys. Rev. Lett.} \textbf{105} (2010) \href{http://link.aps.org/doi/10.1103/PhysRevLett.105.031302}{031302}
[arXiv:\href{http://arxiv.org/abs/0905.3168}{0905.3168} [gr-qc]].

\bibitem{Engle:2010kt}
J.~Engle, K.~Noui, and A.~Perez, \textit{Phys. Rev. D} \textbf{82} (2010) \href{http://link.aps.org/doi/10.1103/PhysRevD.82.044050}{044050}
[arXiv:\href{http://arxiv.org/abs/1006.0634}{1006.0634} [gr-qc]].

\bibitem{Engle:2011vf}
J.~Engle, K.~Noui, A.~Perez, and D.~Pranzetti, \textit{JHEP} \textbf{1105} (2011) \href{http://www.springerlink.com/content/g48jnn447636889v/}{016}
[arXiv:\href{http://arxiv.org/abs/1103.2723}{1103.2723} [gr-qc]].

\bibitem{Krasnov:1997yt} K. Krasnov, \textit{Class. Quant. Grav.} \textbf{16} (1999) \href{http://dx.doi.org/10.1088/0264-9381/16/2/018}{563} [arXiv:gr-qc/\href{http://arxiv.org/abs/gr-qc/9710006}{9710006}].

\bibitem{Kaul:1998xv}
R.~K. Kaul and P. Majumdar, \textit{Phys. Lett.} \textbf{B439} (1998) \href{http://dx.doi.org/10.1016/S0370-2693(98)01030-2}{267} [arXiv:gr-qc/\href{http://arxiv.org/abs/gr-qc/9801080}{9801080}].

\bibitem {Kaul:2000kf}
R.~K. Kaul and P. Majumdar, \textit{Phys. Rev. Lett.} \textbf{84} (2000) \href{http://link.aps.org/doi/10.1103/PhysRevLett.84.5255}{5255} [arXiv:gr-qc/\href{http://arxiv.org/abs/gr-qc/0002040}{0002040}].

\bibitem{Kupeli}
D. N. Kupeli, \textit{Geometriae Dedicata} \textbf{23} (1987)
\href{http://dx.doi.org/10.1007/BF00147389}{33}.


\bibitem{AFK} A. Ashtekar, S. Fairhurst, and B. Krishnan,
 \textit{Phys. Rev. D} \textbf{62} (2000) \href{http://link.aps.org/doi/10.1103/PhysRevD.62.104025}{104025} [arXiv:gr-qc/\href{http://arxiv.org/abs/gr-qc/0005083}{0005083}].



\bibitem{Wald} R. M. Wald, \textit{General Relativity}, Chicago University Press, (1984).


\bibitem{LP1} J. Lewandowski and T. Paw{\l}owski, \textit{Clas. Quant. Grav.} \textbf{20} (2003) \href{http://dx.doi.org/10.1088/0264-9381/20/4/303}{587}
   [arXiv:gr-qc/\href{http://arxiv.org/abs/gr-qc/0208032}{0208032}].

\bibitem{LP2} J. Lewandowski and T. Paw{\l}owski, \textit{Clas. Quant. Grav.} \textbf{22} (2005) \href{http://dx.doi.org/10.1088/0264-9381/22/9/007}{1573} [arXiv:gr-qc/\href{http://arxiv.org/abs/gr-qc/0410146}{0410146}].

\bibitem{LP3} M. Korzy\'nski, J. Lewandowski, and T. Paw{\l}owski, \textit{Clas. Quant. Grav.} \textbf{22} (2005) \href{http://dx.doi.org/10.1088/0264-9381/22/11/006}{2001}   [arXiv:gr-qc/\href{http://arxiv.org/abs/gr-qc/0412108}{0412108}].

\bibitem{LP4} J. Lewandowski and T. Paw{\l}owski, \textit{Clas. Quant. Grav.} \textbf{23} (2006) \href{http://dx.doi.org/10.1088/0264-9381/23/20/022}{6031} [arXiv:gr-qc/\href{http://arxiv.org/abs/gr-qc/0605026}{0605026}].



\bibitem{ABF}
A. Ashtekar, C. Beetle, and S. Fairhurst, \textit{Class. Quant. Grav.} \textbf{16} (1999) \href{http://dx.doi.org/10.1088/0264-9381/16/2/027}{L1}
  [arXiv:gr-qc/\href{http://arxiv.org/abs/gr-qc/9812065}{9812065}].

\bibitem{ABL1} A. Ashtekar, C. Beetle, and J. Lewandowski, \textit{Phys.Rev.} \textbf{D64} (2001) \href{http://link.aps.org/doi/10.1103/PhysRevD.64.044016}{044016}   [arXiv:gr-qc/\href{http://arxiv.org/abs/gr-qc/0103026}{0103026}].


\bibitem{ABE} A. Ashtekar, J. Engle, and C. Van Den Broeck, \textit{Class. Quant. Grav.}
\textbf{22} (2005) \href{http://dx.doi.org/10.1088/0264-9381/22/4/L02}{L27}  [arXiv:gr-qc/\href{http://arxiv.org/abs/gr-qc/0412003}{0412003}].

\bibitem{BE} C. Beetle, and J. Engle, \textit{Class. Quant. Grav.} \textbf{27} (2010) \href{http://dx.doi.org/10.1088/0264-9381/27/23/235024}{235024}
[arXiv:\href{http://arxiv.org/abs/1007.2768}{1007.2768}[gr-qc]].

\bibitem{Basu:2009cw}
R. Basu, R. K. Kaul, and P. Majumdar, \textit{Phys. Rev. D} \textbf{82} (2010) \href{http://link.aps.org/doi/10.1103/PhysRevD.82.024007}{024007}
[arXiv:\href{http://xxx.unizar.es/abs/0907.0846}{0907.0846} [gr-qc]].

\bibitem{FernandoBarbero:2009ai}
 J.~F.~Barbero~G., J.~Lewandowski,  and E.~J.~S.~Villase\~nor,
\textit{ Phys. Rev. D} \textbf{80} (2009) \href{http://link.aps.org/doi/10.1103/PhysRevD.80.044016}{044016}
[arXiv:\href{http://arxiv.org/abs/0905.3465}{0905.3465} [gr-qc]].


\bibitem{Griffiths}
R.~B.~Griffiths, \textit{J. Math. Phys.} \textbf{6} (1965) \href{http://link.aip.org/link/doi/10.1063/1.1704681}{1447}.

\bibitem{Alej}
E.~Frodden, A.~Ghosh and A.~Perez, \textit{A local first law for black hole thermodynamics} [arXiv:\href{http://arxiv.org/abs/1110.4055}{1110.4055} [gr-qc]].

\bibitem{Livine:2005mw}
E.~R.~Livine and D.~R.~Terno, \textit{Nucl. Phys. B} \textbf{741} (2006) \href{http://www.sciencedirect.com/science/article/pii/S0550321306001295}{131-161}
[arXiv:gr-qc/\href{http://arxiv.org/abs/gr-qc/0508085}{0508085}].

\bibitem{Ghosh:2006ph} A. Ghosh and P. Mitra, \textit{Phys. Rev. D} \textbf{74} (2006) \href{http://link.aps.org/doi/10.1103/PhysRevD.74.064026}{064026} [arXiv:hep-th/\href{http://arxiv.org/abs/hep-th/0605125}{0605125}].

\bibitem{Bianchi:2010qd}
E.~Bianchi, \textit{Class. Quant. Grav.} \textbf{28} (2011) \href{http://iopscience.iop.org/0264-9381/28/11/114006/}{114006}
[arXiv:\href{http://arxiv.org/abs/1011.5628}{1011.5628} [gr-qc]].

\bibitem{G.:2008mj}
J.~F.~Barbero~G. and E.~J.~S.~Villase\~nor, \textit{Class. Quant. Grav.} \textbf{26} (2009) \href{http://iopscience.iop.org/0264-9381/26/3/035017/}{035017} [arXiv:\href{http://arxiv.org/abs/0810.1599}{0810.1599} [gr-qc]].

\bibitem{Agullo:2009eq}
I.~Agullo, J.~F.~Barbero~G.,  E.~F.~Borja, J.~Diaz-Polo, and E.~J.~S.~Villase\~nor, \textit{Phys. Rev. D} \textbf{80} (2009) \href{http://link.aps.org/doi/10.1103/PhysRevD.80.084006}{084006}
[arXiv:\href{http://arxiv.org/abs/0906.4529}{0906.4529} [gr-qc]].

\bibitem{Carlip:2000nv}
S. Carlip, \textit{Class. Quant. Grav.} \textbf{17} (2000) \href{http://dx.doi.org/10.1088/0264-9381/17/20/302}{4175}
[arXiv: gr-qc/\href{http://arxiv.org/abs/gr-qc/0005017}{0005017}].

\bibitem{Barrau:2011md} A. Barrau, X. Cao, J. Diaz-Polo, J. Grain, and T. Cailliteau, \textit{Phys. Rev. Lett.} \textbf{107} (2011)   \href{http://link.aps.org/doi/10.1103/PhysRevLett.107.251301}{251301} [arXiv:\href{http://de.arxiv.org/abs/1109.4239}{1109.4239} [gr-qc]].

\bibitem{Giesel:2007wn} K. Giesel and T. Thiemann, \textit{Class. Quant. Grav.} \textbf{27} (2010) \href{http://dx.doi.org/10.1088/0264-9381/27/17/175009}{175009} [arXiv:\href{http://arxiv.org/abs/0711.0119}{0711.0119} [gr-qc]].

\bibitem{Domagala:2010bm} M. Domagala, K. Giesel, K. Kaminski, and J. Lewandowski, \textit{Phys. Rev. D} \textbf{82} (2010) \href{http://link.aps.org/doi/10.1103/PhysRevD.82.104038}{104038} [arXiv:\href{http://arxiv.org/abs/1009.2445}{1009.2445} [gr-qc]].




\end{thebibliography}
\end{document}